\documentclass[12pt]{spieman}  
\usepackage{amsmath,amsfonts,amssymb}
\usepackage{graphicx}
\usepackage{setspace}
\usepackage{tocloft}
\usepackage{ulem}
\usepackage{xcolor}
\usepackage{makecell}
\usepackage{subcaption}
\usepackage{hyperref}
\usepackage{comment}

\title{Design and development of Fabry–Pérot based wavelength calibration system for PARAS-2 spectrograph}

\author[a,b,*]{Shubhendra Nath Das}
\author[a]{Kapil Kumar Bhardwaj}
\author[a]{Abhijit Chakraborty}
\author[a]{Kevikumar A. Lad}
\author[a]{J.S.S.V Prasad Neelam}
\author[a]{Rishikesh Sharma}
\author[a]{Nikitha Jitendran}
\author[a]{Vishal Joshi}

\affil[a]{Astronomy and Astrophysics Division, Physical Research Laboratory, Ahmedabad, Gujarat 380009, India}
\affil[b]{Physics Department, Indian Institute of Technology Gandhinagar, Gandhinagar, Gujarat 382355, India}

\cftpagenumbersoff{figure}
\cftpagenumbersoff{table} 

\begin{document} 
\maketitle

\begin{abstract}
Precise wavelength calibration is essential for high-precision radial velocity (RV) spectrographs, necessitating a stable calibrator that provides a dense grid of uniformly spaced lines to accurately determine stellar line positions and monitor instrumental drifts. In this work, we present the development of a cost-effective Fabry–Pérot (FP) etalon-based wavelength calibrator designed to overcome the limitations of conventional sources such as hollow cathode lamps (HCLs) and iodine cells. This FP calibrator, combined with a Xenon (Xe) arc lamp assembly, has been integrated with the PARAS-2 spectrograph on the PRL 2.5m telescope at Mount Abu Observatory. Operated under controlled temperature and pressure conditions, the system generates a dense, comb-like spectrum covering 62 echelle orders with more than 10,000 well-defined and stable spectral lines, enabling precise measurement of instrumental drift. Initial results show that the free spectral range (FSR) varies from 0.16 \AA~near 4000 \AA~to 0.49 \AA~ near 7000 \AA, with a value of 0.3 \AA~around the central wavelength of 5500 \AA~. The estimated finesse ranges from 9 near 4000 \AA~to 19 near 6900 \AA, with an approximate value of 17 at 5500 \AA. The temperature and pressure stability tests demonstrate RMS variations of $0.002 ^\circ\mathrm{C}$ and $5\times10^{-4}$ mbar, respectively. Based on these values, the theoretical stability of the FP wavelength calibrator is estimated to be within 10 cm/s, establishing it as a reliable alternative to Laser Frequency Combs (LFCs) for high-resolution spectroscopic calibration. We present an initial assessment of the radial-velocity stability of the Fabry–Pérot calibrator, yielding 40-70 cm/s of relative drifts, which are up for further investigations. The observed excess over the theoretically estimated limit is likely attributable to instabilities arising from arc wandering in the xenon arc lamp.
\end{abstract}


{\noindent \footnotesize\textbf{*}Shubhendra Nath Das,  \linkable{shubhendranathdas@gmail.com} }

\begin{spacing}{1}   

\section{Introduction}\label{sec1}

In the field of exoplanet research, the search for smaller planets like sub-Neptune and Super-Earth
has been the focal point of interest \cite{Jeffers_2025_small_planet,Hojjatpanah_2019_small_planet}. The various exoplanet detection techniques, such as transit, RV, astrometry, microlensing, have evolved significantly in the past decade. Despite that, the detection of smaller exoplanets still remains difficult due to several on-sky and off-sky limitations \cite{Oshagh_2018_on-sky_off-sky_limits}. To detect such planets with doppler spectroscopy, an RV precision of $\leq$ $sub~m/s$ is required. These spectrographs uses the simultaneous referencing method to precisely track the instrumental drifts. In this method, the calibration fiber is fed by a stable wavelength calibrator  \cite{Mayor-2003} while acquiring the star's spectra through star fiber. Afterwards, the measured drifts from the calibrator's data is subtracted from the stellar RVs. The poor wavelength calibration of these spectrographs can restrict the RV precision by improper measurement of either the instrumental drifts or wavelengths of the stellar spectral lines. \\
There are several wavelength calibrators available for simultaneous referencing. Among them, emission lamps (Thorium Argon (ThAr) \cite{Lovis_2007_thar},Uranium Argon (UAr) \cite{Sharma_2021_uar}, Uranium Neon (UNe) HCLs etc), Iodine cells \cite{Marcy_1992_Iodine_cell}, Laser frequency comb (LFC)\cite{Ycas_2012_lfc,Murphy_2007_lfc}, and Fabry–Pérot etalon (FP)-based wavelength calibrators \cite{wildi_2010_fp,schafer_2012_fp,cersullo_2017_fp_spirou,banyal_2017_fp,tanya_2018_fp,Cersullo_2019_fp_harps,Tang_2021_fiber-fp} are popular choices. Emission lamps are the traditional choices as wavelength calibration of RV spectrographs. UAr \cite{Sharma_2021_uar, Baliwal_2024_toi6651, Abhijit_2024_bina_uar} or ThAr HCLs\cite{Lovis_pepe_2007_ThAr-linelist} are the most widely used because of their simplicity, easy handling and less maintenance. Currently, here at Physical Research Laboratory (PRL), a UAr HCL is used as the wavelength calibrator with the PRL Advanced Radial-velocity Abu-sky Search-2 (PARAS-2). PARAS-2 is a high resolution fiber-fed spectrograph with a spectral resolution (R = $\dfrac{\lambda}{\delta \lambda}$) of 110,000 at 5500 \AA~ with a working wavelength range from 4000 \AA~ to 7000 \AA, attached with PRL 2.5m telescope at Mount Abu Observatory \cite{Abhijit_2018_paras2,Abhijit_2024_bina_uar}. Although HCLs have proven to be a very dependable source of wavelength calibration \cite{Baliwal_2024_toi6651,Baliwal_2025_toi6038,Khandelwal_2022,Khandelwal_2023}, there are some disadvantages \cite{Kerber_2008,Cersullo_2019_fp_harps,hobson_2021,Reiners_2024} that pose constraints in our way of achieving required RV precision. These HCLs spectra are non-uniformly distributed over an echelle order, and they also contain blended lines. It can have line position measurement uncertainties up to $10 ~m/s$ \cite{Lovis_pepe_2007_ThAr-linelist}.  Apart from that, during the simultaneous observations with the science target, bright transitions from filled buffer gas~$i.e.$~Argon/Neon can contaminate the nearby region of the star's spectra by cross-talking and bleeding. With time, the HCLs tend to become inefficient by producing poor spectra. Therefore, a better wavelength calibrator is required to achieve the required RV precision. Based on these observations, the desired properties for an ideal wavelength calibrator are listed below
\cite{cersullo_2017_fp_spirou}:
\begin{itemize}
    \item Large number of lines
    \item Well-resolved lines
    \item Uniformly distributed over an echelle order
    \item Good SNR for each line
    \item Coverage of full working wavelength range
    \item Stable device over a long period of time
    \item Long lifetime
\end{itemize}
Calibrators, such as LFC and FP etalon wavelength calibrators, possess these characteristics and closely adhere to these properties. Though LFC is a very good option to be used as a wavelength calibrator, its high cost and operational difficulties limit its accessibility to groups with a limited budget. With a similar type of comb-like spectra, the FP wavelength calibrator provides a large number of high SNR lines at an affordable cost and with comparatively fewer operational difficulties. Therefore, here at PRL, we planned to implement the FP wavelength calibration system with PARAS-2, which will be used to enhance the instrumental RV measurement capabilities. In this paper, we discuss the complete design and development of the FP wavelength calibration system. Section 2 discusses the fundamentals of an FP etalon, while Section 3 focuses on the design of the FP system. We discuss the pressure and temperature stability requirements in Section 4 and show initial results from the FP calibration system along with its stability in Section 5. We summarize our work and possible future improvements in Section 6.

\section{Fundamental concepts of FP etalon-based wavelength calibrator}
An FP etalon is an optical element made up of two flat mirrors (in the case of a plano-plano Fabry-Pérot, although other configurations are also possible) spaced by a fixed distance. Both mirrors are slightly wedged to avoid reflections from their outer surfaces and also coated with anti-reflective (AR) coatings. This operates on the basis of the principle of multiple-beam interferometry. It divides the amplitude of the incident light ray (division of amplitude interference) into multiple beams due to its inner highly reflective mirror surfaces. When illuminated with polychromatic light, the FP etalon acts as a spectral filter, allowing only specific wavelengths that satisfy the condition of constructive interference. The key parameters influencing this condition include the spacer length (distance between the two mirrors) ($d$), the refractive index of the medium between the mirrors ($n$), and the angle of incidence ($\psi$)(or angle of refraction ($\theta$)) (refer to Figure \ref{fig:fp-schematic}). This interference condition is expressed as \cite{hecht_2002_optics_book}: 
\begin{equation} 
2nd\cos \theta = m \lambda \label{eq:fp-interference}
\end{equation} 
Here $m$ is the interference order of the FP's spectra, and $\lambda$ is the associated central wavelength.

\begin{figure}[b]
\centering
\includegraphics[width=0.7\linewidth]{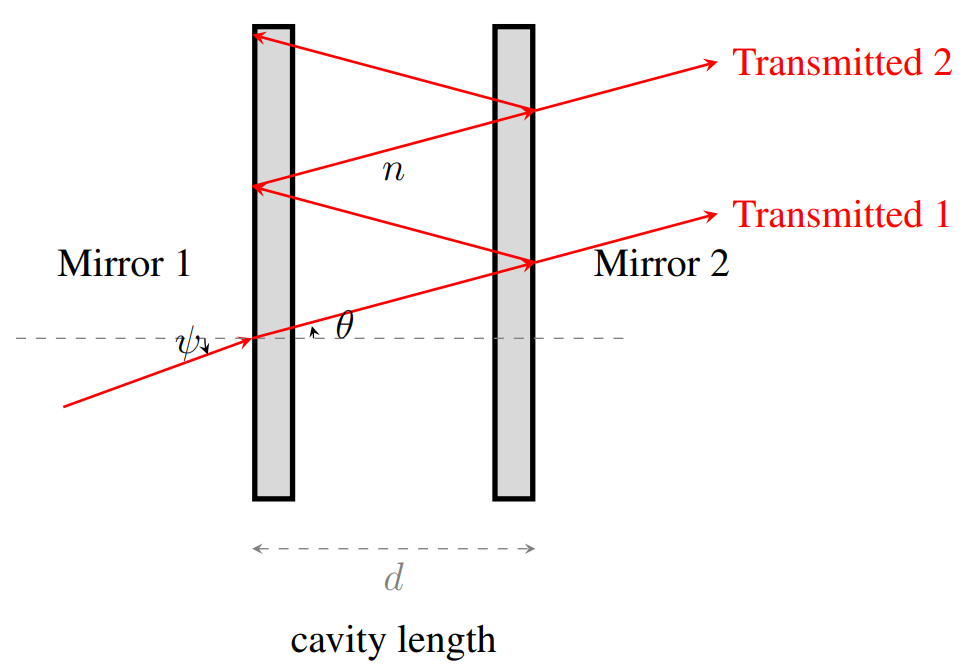}
\caption{Schematic of the FP cavity showing the incident angle ($\psi$) and refraction angle ($\theta$). The \textbf{Transmitted 1} and \textbf{2} beams, under constructive interference condition from Equation~\ref{eq:fp-interference}, produce observable emission peaks.}
\label{fig:fp-transmission function}
\end{figure}

The transmission function at the output of a FP etalon can be obtained by summing the intensities of all the transmitted rays. This derivation is available in most standard optics textbooks \cite{hecht_2002_optics_book,ghatak_2009_optics_book}. The resulting transmission function is known as the Airy function and is given by \cite{hecht_2002_optics_book}, 
\begin{equation} 
I = \dfrac{I_0}{1+F\sin^2(\frac{\delta}{2})} 
\label{eq:fp-transmission_function} 
\end{equation} 
In this expression, $\delta$ describes the path difference between two consecutive transmitted rays, $I_0$ is the intensity at the peak, and $F$ denotes the finesse, which is related to the Full-Width at Half Maximum (FWHM) of each spectral line produced by the FP etalon. Under ideal conditions, finesse is only determined by the reflectivity of the mirror coatings, as expressed in Equation (3). However, in practical applications, finesse also depends on other factors such as parallelism of the FP mirrors, surface smoothness of the high-reflective coating, size of the input source (size of the input fiber in most of the cases) e.t.c as mentioned in Cersullo (2017) \cite{cersullo_2017_fp_spirou}. In Equation \ref{eq:fp-transmission_function}, finesse is expressed solely as a function of reflectivity (R) as given in \cite{hecht_2002_optics_book}: 
\begin{equation}
F = \dfrac{\pi \sqrt{R}}{1-R} \label{eq:fp-finess} 
\end{equation} 
The finesse factor influences both the FWHM of the spectral lines and the background level, leading to light loss. As shown in Figure \ref{fig:fp-transmission function}, a decrease in the mirror reflectivity of the FP's coating from 90\% to 30\% results in a noticeable increase in both the FWHM and the background level of each spectral line relative to its peak intensity. Transmission also depends on the phase difference ($\delta$) between successive transmitted rays. This phase difference can be derived as \cite{hecht_2002_optics_book}: 
\begin{equation} 
\delta = \dfrac{4\pi n d \cos \theta}{\lambda}
\label{eq:fp-phase_difference} 
\end{equation}

\begin{figure}[ht]
\centering
\includegraphics[width=0.9\linewidth]{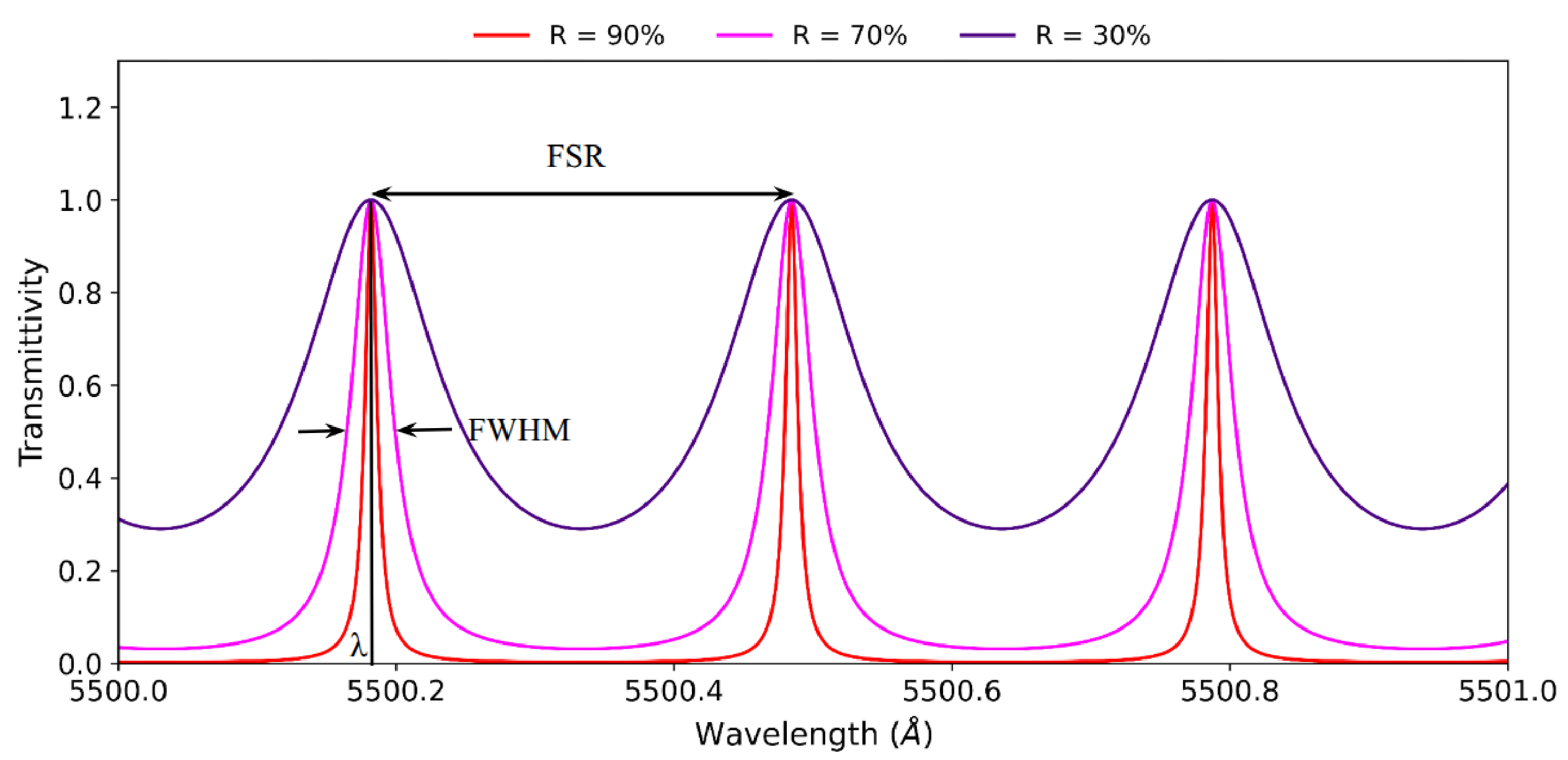}
\caption{Simulated transmission function of FP etalon for polychromatic light incident upon the device. Transmission functions are drawn for different reflective mirror surfaces.}
\label{fig:fp-transmission function}
\end{figure}

In the transmission profile of the FP etalon, the separation between two consecutive spectral lines is known as the free spectral range (FSR). This FSR can be written as a function of wavelength and the separation of the FP's mirrors under the approximation of parallel beam incident ($\psi$=$\theta$=0) as \cite{hecht_2002_optics_book},
\begin{equation}
    \text{FSR} = \Delta \lambda = \dfrac{\lambda^2}{2nd}
    \label{eq:fsr}
\end{equation}
In this context of FSR, we can also define finesse as the ratio between the FSR and the FWHM \cite{Jacquinot1960}.

\section{Design of the FP etalon wavelength calibrator}

The FP etalon wavelength calibration system can be thought of as two main subsystems: (1) a white light source for illuminating the Fabry-Pérot and (2) the FP optical assembly itself. Unlike emission lamps, the FP calibrator is a passive device that requires illumination with uniform-intensity white light over the operating wavelength range. In our case, this range spans from 4000 \AA~to 7000 \AA. Several light sources provide such uniform intensity, featureless flux over the required wavelength range, including laser-driven light sources (LDLS) \cite{Schmidt_2022_chromatic_drift_fp_ldls}, supercontinuum sources \cite{Schwab_2015_fp_superK}, intensity-stabilized broadband fiber-coupled LEDs \cite{tanya_2018_fp}, Xe arc lamps \cite{wildi_2010_fp} and a few others. Among these, the Xe arc lamp stands out as a cost-effective choice, offering a relatively uniform continuum flux across the visible spectrum from 4000 to 7000 \AA~.

\begin{figure}[t]
	\centering
	\includegraphics[width=0.99\linewidth]{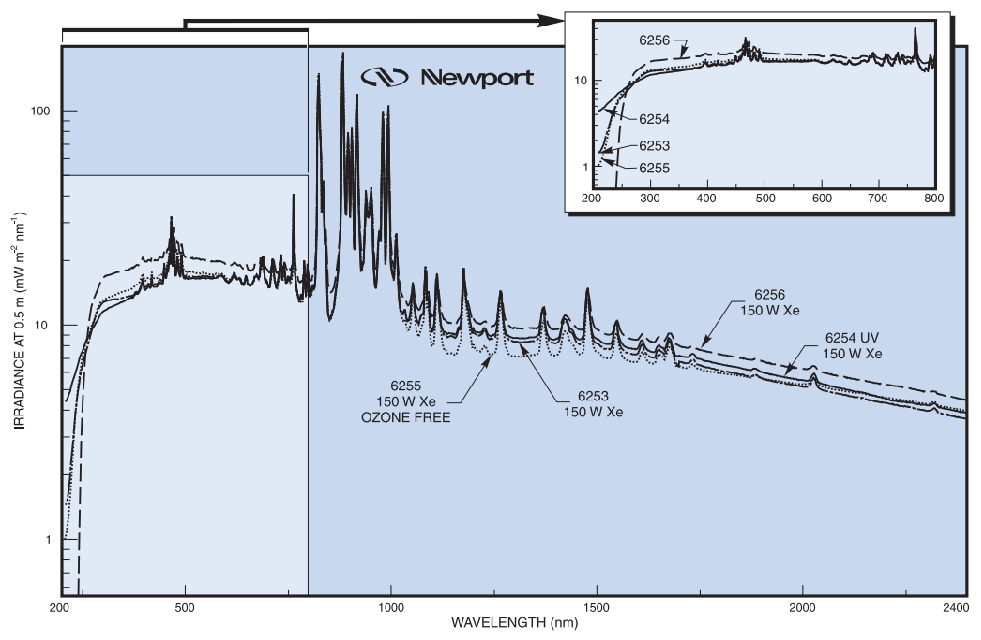}
	\caption{The spectra of the Xe arc lamp taken from the Newport website (\url{https://www.newport.com/f/xenon-arc-lamps}). The plot in the upper-right side is showing the zoomed part from 200 nm to 800 nm . A nearly uniform flux is seen in the wavelength range of 400 nm to 700 nm. There is also a good amount of flux in the NIR and IR regions. (N.B: Here  the wavelength scale is used in nm unit)}
	\label{fig: Xe arc lamp spectra}
\end{figure}

We can see the spectrum of the Xe arc lamp in Figure \ref{fig: Xe arc lamp spectra}, which demonstrates a nearly uniform flux in the aforementioned wavelength range.

\subsection{Design of the white light source (Xenon arc) assembly }
At PRL, we have designed and developed a custom Xe arc lamp housing with the primary goal of maximising light collection while effectively filtering out the IR excess present in the lamp's spectrum. The key optical components in this system include a collimator lens, a focuser lens, and an IR-blocking filter alongside a 150-watt Xe arc lamp. We have employed a combination of achromatic doublet and triplet for the collimation and focusing stages.
While alternative designs using elliptical or spherical mirrors exist, we opted for a lens-based configuration due to its simplicity in fabrication, easy optical alignment and spacious layout, which facilitates better heat dissipation from the Xe arc lamp.

We have used off the shelf available optical components. As a collimator, we use the EO 32-886 achromatic doublet from Edmund Optics\footnote[1]{https://www.edmundoptics.in/}~(diameter = 50.00 mm, EFL = 150.00 mm), while the PAC076 achromatic triplet from Newport \footnote[2]{https://www.https://www.newport.com/}~ (diameter = 38.10 mm, EFL = 125.00 mm) is used for focusing. The illumination source is a 150-watt Xe arc lamp from Techinstro Industries \footnote[3]{https://www.techinstro.com/xenon-arc-lamp/}, which delivers approximately 3200 lumens of total luminous flux with an arc size of around 2.6 mm. These lamps are rated for stable performance over a lifespan of more than 750 hours.

\begin{figure}[t]
	\centering
	\includegraphics[width=0.7\linewidth]{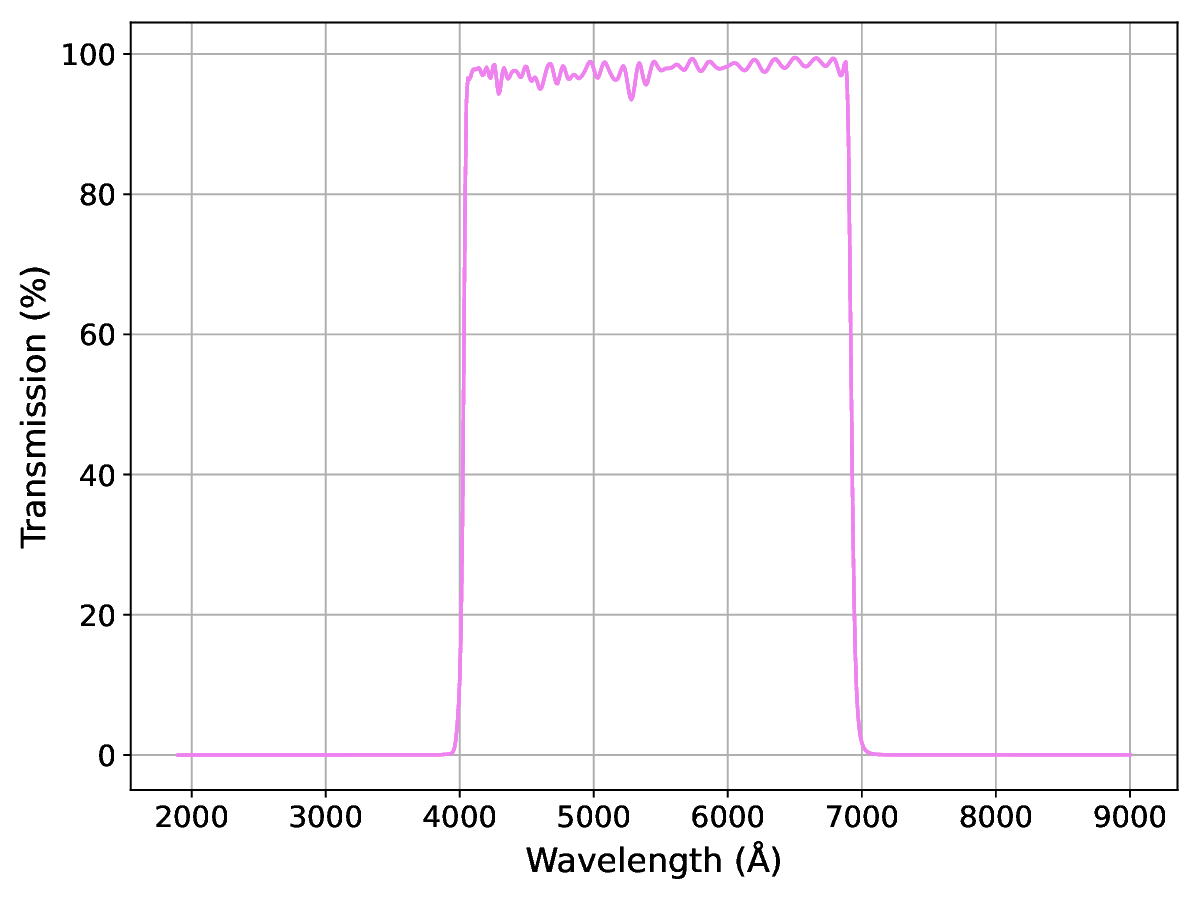}
	\caption{The transmission response function of the bandpass filter used in the Xe arc lamp housing (EO 89-799) showing $\sim$90\% transmission in 4000-7000 \AA~ wavelength range. This high transmission across the visible spectrum makes it well-suited for our application by effectively blocking out-of-band radiation, particularly in the IR region.}
	\label{fig:bandpassfilter}
\end{figure}

The FP wavelength calibrator is designed to work between the wavelength range of 4000 \AA~to 7000 \AA. Therefore, we are using a band-pass filter (EO 89-799 from Edmund Optics) to only allow the required wavelength range from the Xe arc lamp's spectra and block excess IR radiation. The transmission response function of the filter is shown in Figure \ref{fig:bandpassfilter}. It can be seen that the filter has an almost uniform response across the visible band, accompanied by a sharp decline in the transmission function in the region $\leq$4000~\AA~and $\geq$7000~\AA. This IR suppression is very important, as excess infrared radiation can lead to heating of the Fabry-Pérot's soft optical coatings, potentially degrading their performance by showing chromatic variation \cite{Terrian_2021}.

\begin{figure}[t]
	\centering
	\includegraphics[width=0.99\linewidth]{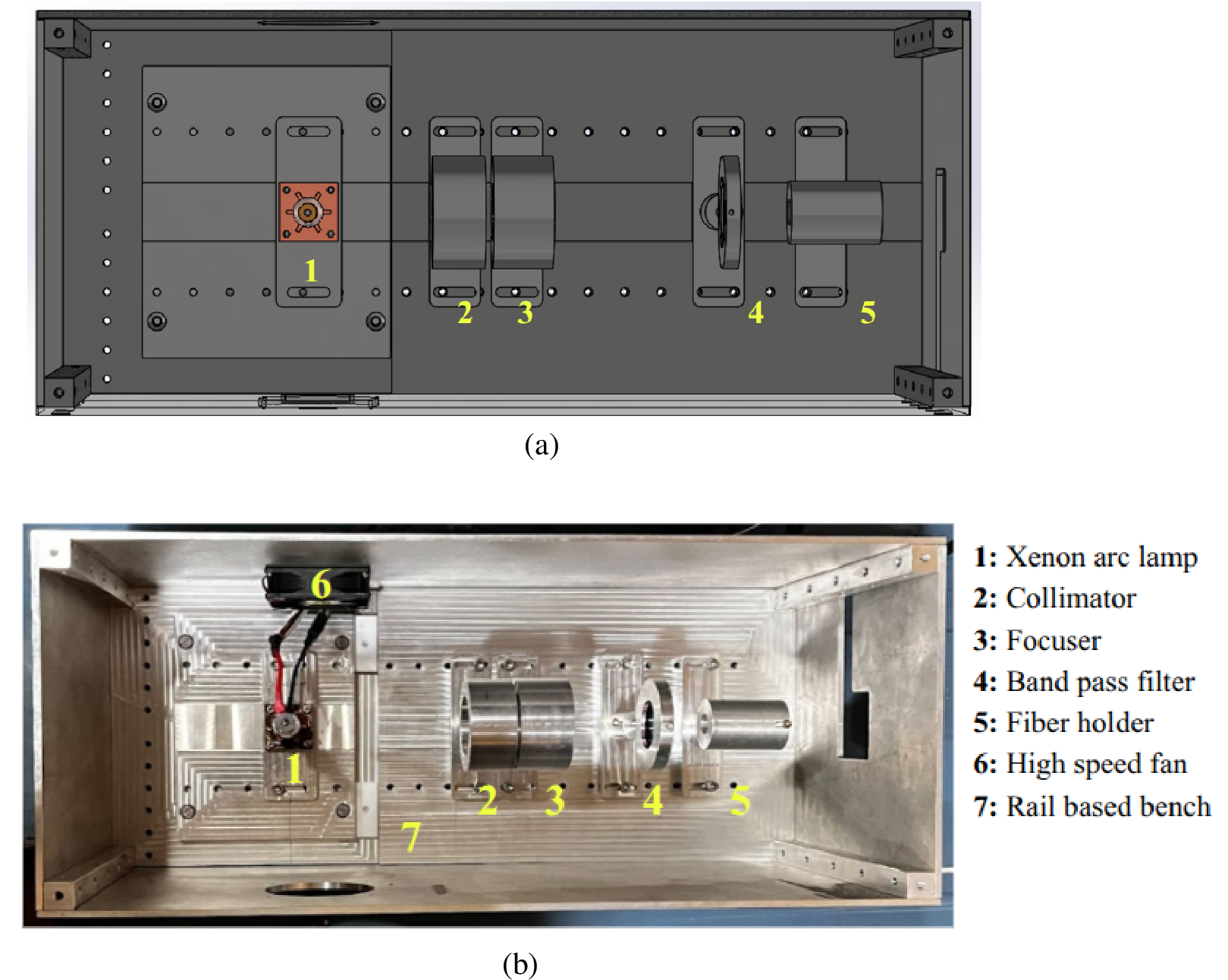}
	\caption{(a)Top view of the SolidWorks design of the Xe arc lamp assembly. (b) The top view of the fabricated Xe arc lamp assembly, captured prior to enclosure, reveals the internal layout and alignment of the optical components. This view clearly displays (from the left) the placement of the Xe arc lamp, collimator, focuser, bandpass filter, and fiber holder along the optical rail. It also highlights the spacious internal design, which facilitates heat dissipation and allows ease of access for adjustments and maintenance.
 }
	\label{fig: real Xe arc lamp top view}
\end{figure}

The enclosure of the Xe arc lamp assembly is fabricated using Al-6061, chosen for its light weight properties, ease of handling, commercial availability, and machinability. To manage the thermal load and ozone generation from the UV component of the Xe arc spectrum, we have integrated two 12 V high-speed fans. These fans help in both heat dissipation and in venting ozone gas effectively. The entire setup is placed in a well-ventilated room, isolated from both the spectrograph and telescope rooms, to ensure thermal and illumination isolation.  All optical and fiber holders shown in Figure \ref{fig: real Xe arc lamp top view} are mounted on a rail-based platform, allowing precise adjustments along the optical axis, as well as flexibility to adjust their height, angle, and position as required for optimal alignment.
The light from the Xe arc lamp is precisely focused onto a 50 $\mu$m core octagonal optical fiber with a numerical aperture (NA) of 0.22, which transfers the light to the FP system. The use of a 50 $\mu$m core octagonal fiber is motivated by the need to enhance the finesse and improve RV stability, as its better scrambling properties can effectively mitigate the arc wandering effects of the Xe arc lamp.\footnote{https://zeiss-campus.magnet.fsu.edu/tutorials/arclampinstability/indexflash.html}

\subsection{Design of the FP wavelength calibrator}
The details of the FP etalon used in this work are listed in Table \ref{table:fp:fp}, supplied by ICOS, UK. Figure \ref{fig:p-2_fpe} (a) shows the FP etalon mounted in its black anodized holder. The transmission function of the FP etalon coating is shown in Figure \ref{fig:p-2_fpe} (b), exhibiting an approximately constant transmission of 5\% across the working wavelength range. This transmission behavior is consistent with the 95\% reflective coating value specified in Table \ref{table:fp:fp}.
Motivated by these etalon characteristics, we now describe the optical design of the FP wavelength calibrator and the key considerations that guided the selection of its components.
\subsubsection{Design considerations}\label{sec:design consideration}
\begin{figure}[b]
	\centering
	\includegraphics[width=0.9\linewidth]{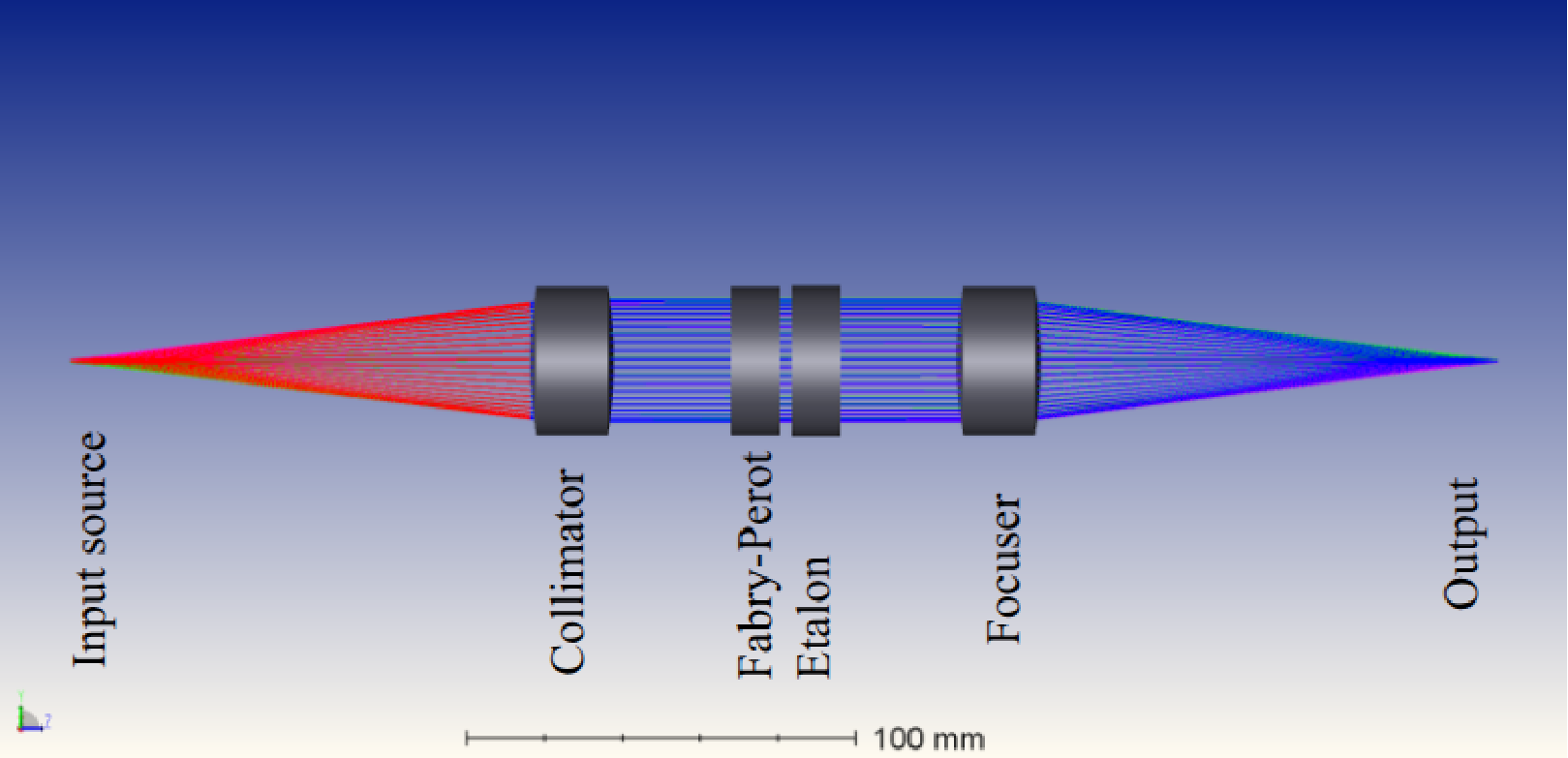}
	\caption{The Zemax optical design of the FP wavelength calibration system illustrates the arrangement of key optical components from left to right: a triplet lens acting as the collimator, followed by the FP etalon, and finally another triplet lens serving as the focuser.}
	\label{fig:fp optical design}
\end{figure}

The optical design of the FP wavelength calibrator contains a collimator, an etalon, and a focuser (see Figure~\ref{fig:fp optical design}). Depending on design preferences and space constraints, either mirrors \cite{cersullo_2017_fp_spirou} or lenses \cite{wildi_2010_fp} can be used for collimation and focusing. In our setup, we have opted for achromatic triplet lenses (PAC076 from Newport) for both collimation and focus in order to minimize chromatic aberration across the operating wavelength range.

\begin{table}[]
\caption{Details of the FP etalon supplied by the manufacturer (I C Optical System (ICOS), UK).}
\label{table:fp:fp}
\begin{center}       
\begin{tabular}{|l|l|p{8cm}|} 
\hline
\rule[-1ex]{0pt}{3.5ex} \textbf{Sl. No} & \textbf{Component} & \textbf{Specification}  \\
\hline\hline
\rule[-1ex]{0pt}{3.5ex} 1 & Model & From IC Optical Element \\
\hline
\rule[-1ex]{0pt}{3.5ex} 2 & Spacer length (mm) & 5 \\
\hline
\rule[-1ex]{0pt}{3.5ex} 3 & FSR (\AA) & 0.3 \\
\hline
\rule[-1ex]{0pt}{3.5ex} 4 & Spacer material & Ultra-low expansion material from Corning \\
\hline
\rule[-1ex]{0pt}{3.5ex} 5 & Reflectivity & 95\% \\
\hline
\rule[-1ex]{0pt}{3.5ex} 6 & Parallelism & $\lambda$/50 or better (at 6330 \AA) \\
\hline
\rule[-1ex]{0pt}{3.5ex} 7 & Flatness & $\lambda$/100 (at 6330 \AA) \\
\hline
\rule[-1ex]{0pt}{3.5ex} 8 & Diameter of clear Aperture (mm) & 50.8 \\
\hline
\end{tabular}
\end{center}
\end{table}

As illustrated in Figure~\ref{fig:fp-transmission function}, the spectrum produced by an FP etalon typically exhibits three main features: FSR, FWHM, and the exact central wavelength of the spectral lines. The performance of the FP wavelength calibrator is mainly affected by several key parameters like the FP optics, the diameter of the input and output fibers, and the environmental stability, particularly with respect to pressure and temperature. In this section, we present the key considerations underlying the design choices for various components of the FP wavelength calibrator. These include the selection of FP optics, the diameter of the input fiber, and the maximum permissible decentering of the input source relative to the optical axis of the collimator.

 In our case, the refractive index and the etalon mirror spacing are selected such that the FSR is five times the FWHM of the corresponding spectral line near the middle of the working wavelength range of the PARAS-2. This allows us to precisely characterize the spectral lines as they do not blend with adjacent lines and have the sufficient number of transmission peaks at the same time. PARAS-2 spectrograph's Instrumental Profile (IP) has an FWHM of $\approx$0.05~\AA~~near 5500~\AA~. Thus, the required FSR would be:
\begin{equation}
\text{FSR} \geq 5 \times \text{IP}_\text{PARAS-2} = 0.25~\text{\AA~}
\end{equation}
Under this criterion, we have chosen an FSR of 0.3 \AA~near 5500 \AA~, which corresponds to a spacer length of 5~mm (from Equation \ref{eq:fsr}) for an air-spaced etalon.

\begin{figure}[b]
\centering
\includegraphics[width=0.9\linewidth]{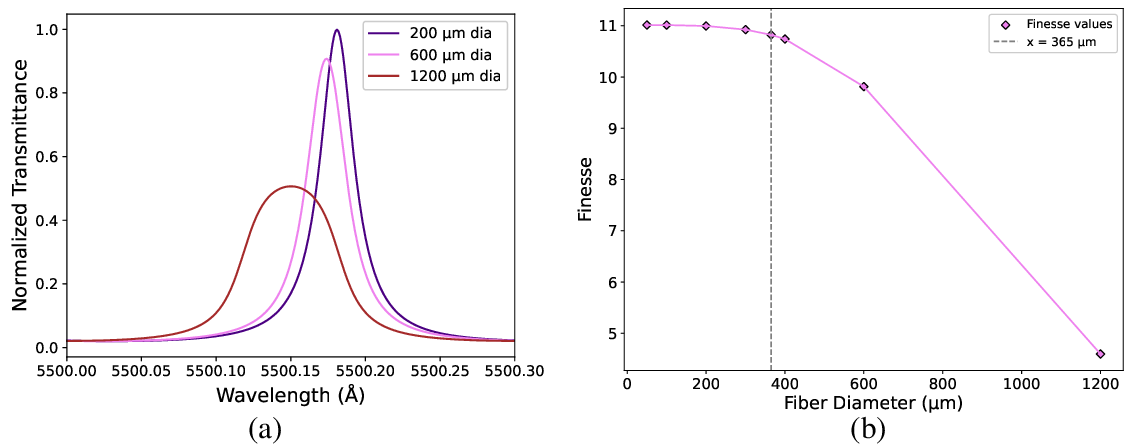}
\caption{(a) The FWHM of each spectral line increases with the increasing diameter of the input fiber, along with a noticeable blue shift in the spectral lines. (b) The finesse value decreases as the input fiber diameter increases, which is evident from the broadening of the spectral lines. The finesse value is estimated by measuring the FWHM around the wavelength 5500 \AA~and considering a FSR of 0.3 \AA. The vertical dashed line in (b) represents the finesse value for the optical fiber size of 365 $\mu$m as input, which agrees with our targeted finesse value of 10.
}
\label{fig:spectral line degradation with fiber dia}
\end{figure}

The second key feature is the FWHM of each spectral line, which is governed by the finesse of the FP system. The finesse affects the amount of light lost after convolution with the IP of the spectrograph. To avoid photon loss, the FWHM of the etalon lines must be smaller than the FWHM of the spectrograph's IP by maintaining an appropriate finesse. Using the method mentioned in Cersullo (2017)\cite{cersullo_2017_fp_spirou}, we have estimated the minimum finesse requirement for our FP wavelength calibrator. With the spectral resolution of 1,10,000 \cite{Abhijit_2024_bina_uar} at the 6900 \AA~ wavelength—and an etalon with a spacer size of 5 mm, the targeted finesse would be $\approx$10. However, as shown in Cersullo (2017)\cite{cersullo_2017_fp_spirou} the effective finesse degrades with the increasing size of input fiber and its decentering from the optical axis. We have simulated these effects for our case as well as shown in Figure~\ref{fig:spectral line degradation with fiber dia} by following the similar procedure mentioned in  Cersullo (2017) \cite{cersullo_2017_fp_spirou}. In this simulation, we aimed to choose the input-fiber diameter for the FP calibrator, given our requirement of achieving a finesse of 10. Although our target finesse is 10, adopting a configuration that yields a higher finesse would further reduce the FWHM of the spectral lines, thereby improving the attainable radial-velocity precision by reducing fundamental radial velocity photon noise. It is evident that with an increasing size of the input fiber, the FWHM increases, and the finesse decreases due to the increment in divergence angle \cite{cersullo_2017_fp_spirou}. 
Therefore we have selected an input fiber of 50 $\mu$m core octagonal fiber, which can gives us our required finesse value  (shown by vertical dashed line in Figure~\ref{fig:spectral line degradation with fiber dia}) as well as the scrambling properties of the octagonal fiber will aid us achieving a better near field and far field illumination at the input side of the FP.

\begin{figure}[b]
\centering
\includegraphics[width=0.9\linewidth]{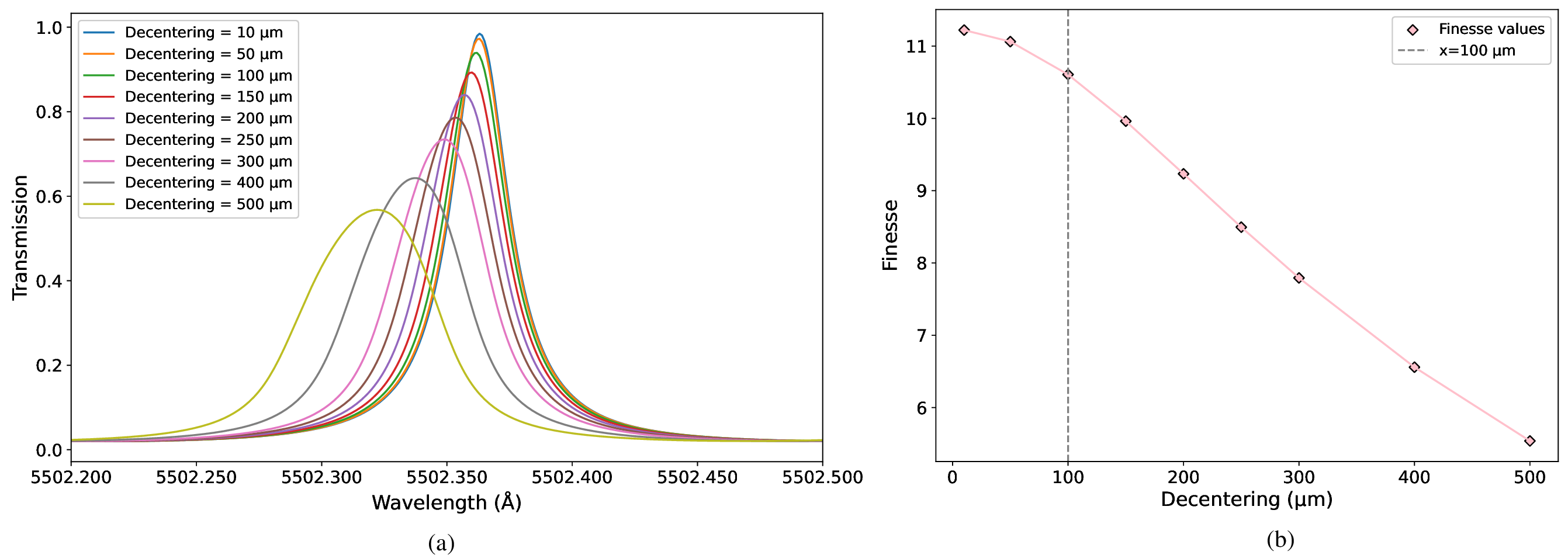}
\caption{(a) Similar to Figure \ref{fig:spectral line degradation with fiber dia}, the spectral lines of the FP exhibit an increase in FWHM along with a blue shift as the distance between the input fiber's center and the optical axis increases. (b) The degradation in finesse with this decentering effect.}
\label{fig:spectral line degradation with decentering}
\end{figure}

De-centering of the optical axis of the FP optical components introduces effects on the FWHM of the spectral lines similar to those caused by increasing the fiber diameter \cite{cersullo_2017_fp_spirou} and degrade the finesse. We simulated these effects as shown in Figure \ref{fig:spectral line degradation with decentering}. Decentering degrades the finesse value more viciously than the increment of the fiber diameter. 

The third important aspect is the central wavelength of each spectral line, which is determined by the refractive index, the angle of incidence, and the mirror spacing, as expressed in Equation~\ref{eq:fp-interference}. It is crucial to maintain a stable refractive index and mirror separation to ensure spectral line stability by controlling operating pressure and temperature. It can be attained by keeping the whole system in a temperature and pressure-controlled environment. The details of the required stability will be discussed in later sections (see \S\ref{sec:stability}).

\subsubsection{Mechanical design of the FP system}
\begin{figure}[t]
    \centering
    \includegraphics[width=0.99\linewidth]{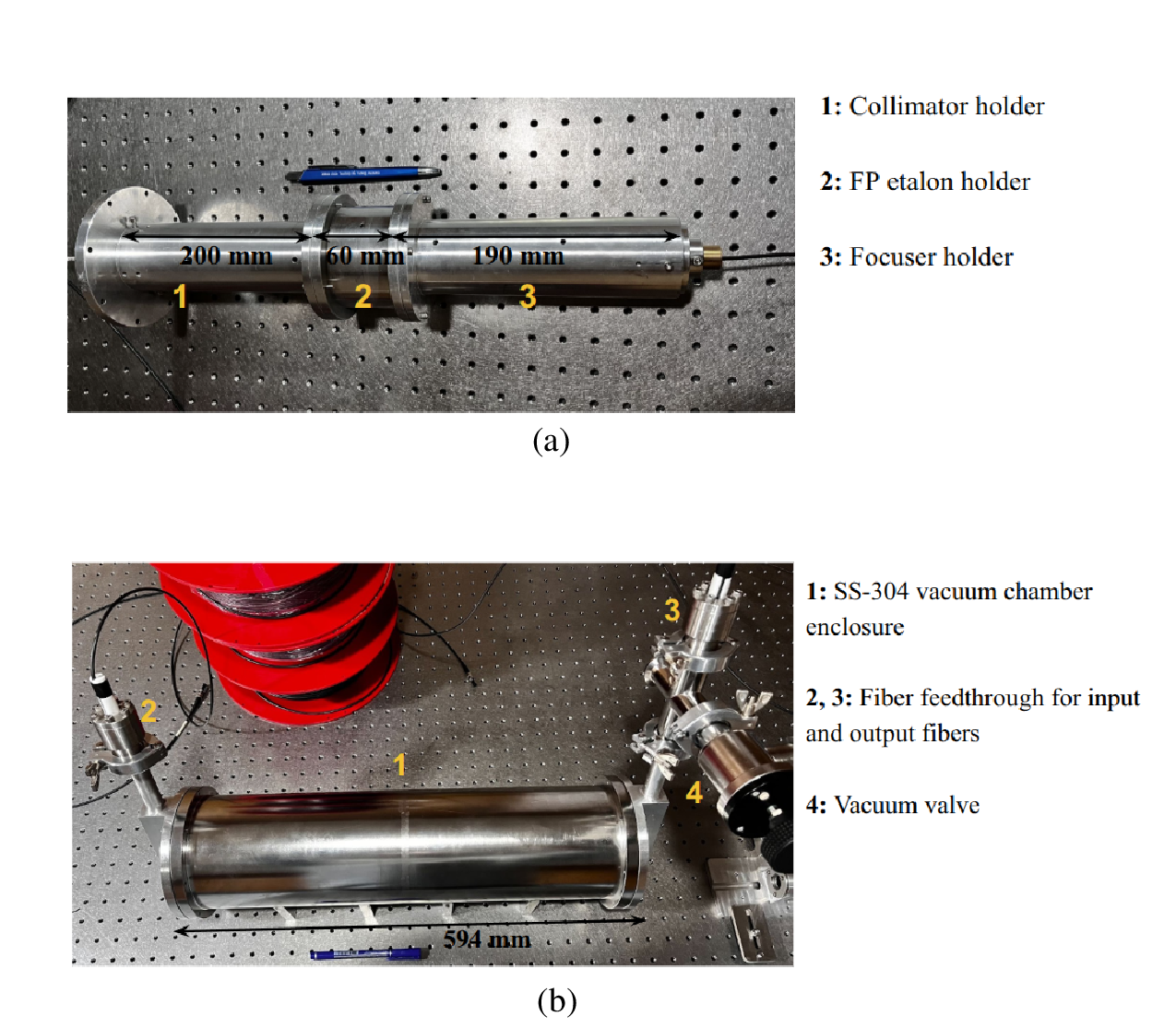}
    \caption{(a)The fabricated aluminium housing (b) Vacuum enclouser for the FP wavelength calibrator }
    \label{fig:fpe real}
\end{figure}

\begin{figure}
    \centering
    \includegraphics[width=0.99\linewidth]{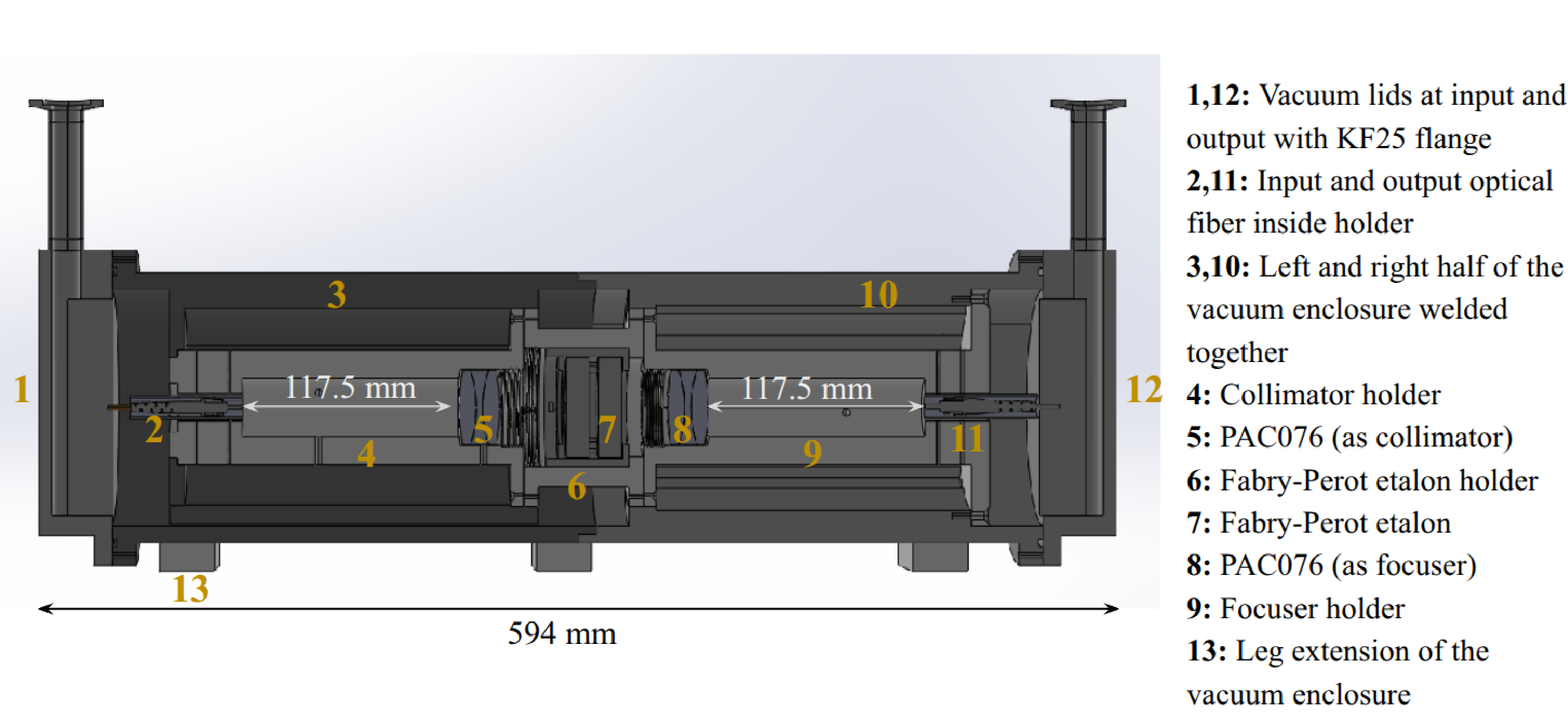}
    \caption{CAD cross-sectional view of the FP system enclosed within the vacuum chamber Figure \ref{fig:fpe real} (lower panel) designed in Solidworks. The two differently shaded sections represent the left and right halves of the vacuum chamber, which are later vacuum-welded along the central plane. The overall optomechanical design exhibits mirror symmetry about the FP etalon, which is centrally located. From left to right, the optical components shown are the collimator lens, the FP etalon, and the focuser lens.}
    \label{fig:fp_cs}
\end{figure}

The mechanical structure of the FP wavelength calibrator primarily consists of two main components namely, the inner tube and the vacuum chamber. The optomechanical holders for the collimator lens, focuser lens, and the FP etalon screwed to each other to form a tube like structure as shown in Figure \ref{fig:fpe real} (upper panel). This tube is then placed inside a vacuum chamber as shown in Figure \ref{fig:fpe real} (lower panel)  and Figure \ref{fig:fp_cs}. The optomechanical holders are fabricated inhouse at PRL workshop. Al-6061 material is chosen for its light weight and relatively low and comparative coefficient of thermal expansion (CTE) to the glass materials of the optical components. The vacuum chamber is made out of SS-304 material. The choice of SS-304 material offers the required structural strength, long-term durability, and vacuum compatibility with a low degassing rate.

The end lids of the vacuum chamber are equipped with KF 25 flanges to connect a vacuum pump at one end and fiber feed through at the other end. A cross-sectional view of the complete system, is provided in Figure \ref{fig:fp_cs}. As shown, the inner tube is supported at one end and screwed at the other end. Additional support is provided in the middle and at the end to minimise flexures over time.

\section {Stability requirement of the FP system}\label{sec:stability} 
 
PARAS-2 is designed to detect sub-meter-per-second radial velocity (RV) signals induced by super-Earth to sub-Neptune-sized exoplanets orbiting Sun-like stars. To reliably measure such small signals, it is critical to minimize the overall error budget. Therefore, our new calibration system has been designed to ensure instrumental stability well below the smallest RV signals detectable by PARAS-2. In particular, the FP calibrator has been engineered to maintain a stability better than 10 cm/s (target stability for the course of 12 hrs (typical night duration for observation)), significantly lower than the detection threshold of the instrument. The two primary environmental factors that can influence the stability of the FP are pressure and temperature by affecting the refractive index of the medium and the spacer length between the FP mirrors. 

\begin{figure}[t]
\centering
\includegraphics[width=0.9\linewidth]{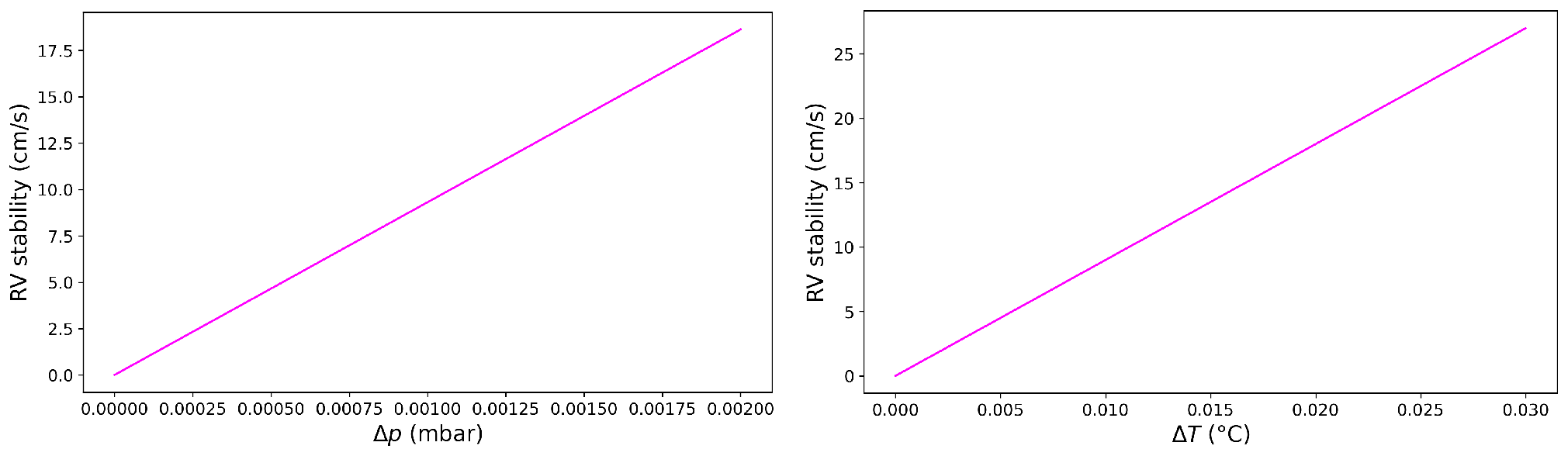}
\caption{Variation in the spectral lines of the FP due to changes in pressure (left panel) and temperature (right panel). Here we have assumed a $p_\text{atm}$ of 870 mbar and CTE of the Corning ULE (spacer material) as $3\times10^{-8} /^\circ\mathrm{C}$.}
\label{fig:fp_temp_variation}
\end{figure}

Variations in pressure can change the refractive index of the medium, and this can shift the positions of the spectral lines. This can be seen from the refractive index dependence of the central wavelengths in Equation \ref{eq:fp-interference}.
Using the relation between refractive index of the medium and pressure as mentioned in \cite{cersullo_2017_fp_spirou},

\begin{equation}
    n = 1 + \dfrac{0.00027\times p}{p_\text{atm}}
\end{equation}
and the relationship between pressure variations and RV shifts via $\dfrac{\Delta v}{c}=\dfrac{\Delta n}{n}$ (where all notations retain their conventional meanings), we get the following relation between the shifts in spectral lines of FP (in terms of RV shifts) with the pressure as:
\begin{equation}
\dfrac{\Delta v}{c} = 2.7\times10^{-4} \dfrac{\Delta p}{p_\text{atm}}\left(\dfrac{0.00027\times p}{p_\text{atm}}+1\right)^{-1}
\end{equation}
The bracketed term on the right-hand side is approximately close to unity, given that the operating pressure ($p$) is much lower than the atmospheric pressure ($p_\text{atm}$). Consequently one can see in the left panel of Figure \ref{fig:fp_temp_variation}, to ensure a RV stability of 10 cm/s, pressure variations must be limited to below $1\times 10^{-3}$ mbar. This estimation assumes an atmospheric pressure of 870 mbar, typical at the PRL Mount Abu Observatory, where the system will be deployed.

The thermal expansion resulting from the temperature variations changes the separation between the etalon mirrors ($d$) as per the equation:
\begin{equation}
    \Delta d = d \alpha \Delta T
\end{equation}
In this equation, $\Delta d$, $\alpha$, and $\Delta T$ are the change in the spacer length, CTE of the spacer material, and the change in temperature, respectively. Using this equation with the Equation \ref{eq:fp-interference}, we can derive the temperature variation dependence on the RV shifts using $\dfrac{\Delta v}{c}=\dfrac{\Delta d}{d} $:
\begin{equation}
    \dfrac{\Delta v}{c}=\alpha \Delta T
    \label{eq:fp_temp}
\end{equation}
Equation \ref{eq:fp_temp} shows that, we need to choose the spacer material which has a very low CTE to achieve the requisite 10 cm/s RV stability. We are using Corning Ultra Low Expansion (ULE) material as the spacer, which has a certified CTE below $3\times 10^{-8} /^\circ\mathrm{C}$. We have shown in the right panel of Figure \ref{fig:fp_temp_variation}, that with the temperature stability of $0.01~^\circ\mathrm{C}$, we can achieve our targeted RV stability of 10 cm/s.

\begin{figure}[b]
\centering
\includegraphics[width=0.9\linewidth]{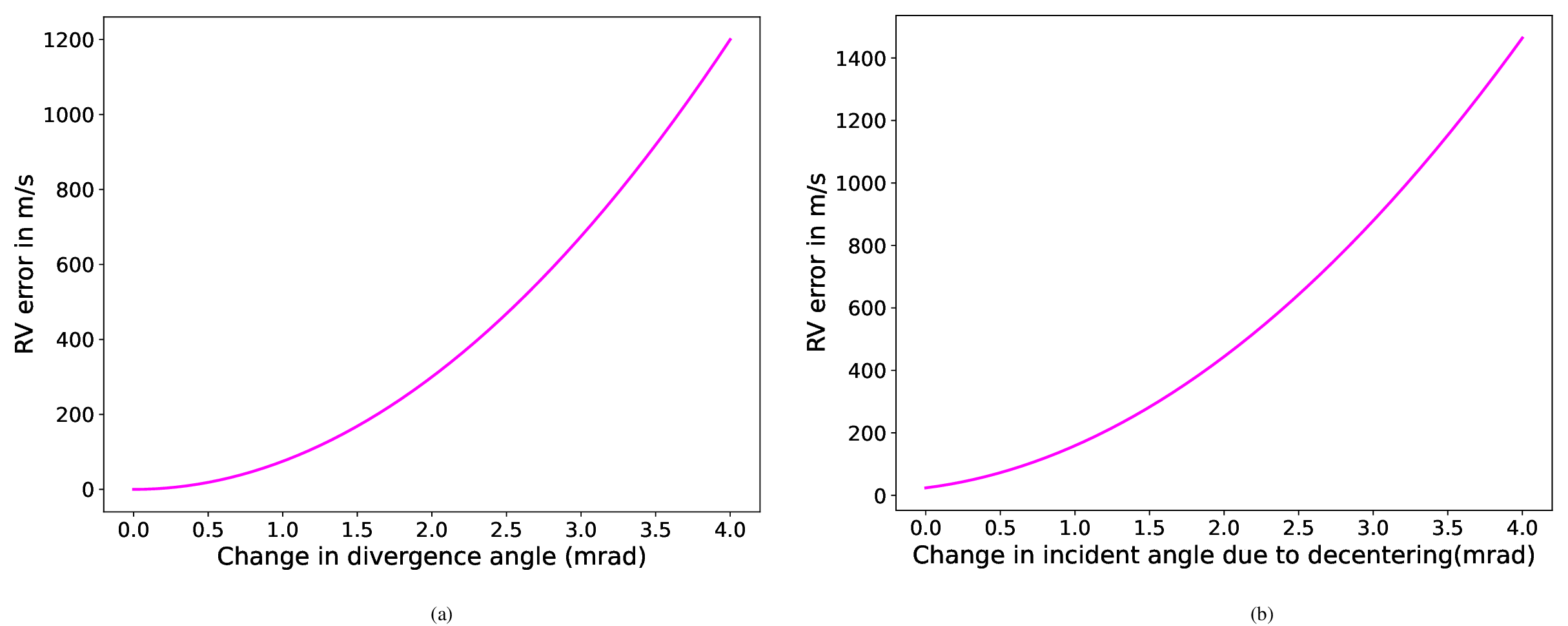}
\caption{RV variation due to (a) change in divergence angle and (b) change in decentering. For the left panel of the figure we have assumed a perfectly centered system. Whereas the right panel considered a 365 \textcolor{blue}{$\mu$m} fiber tip alongside a 125 mm focal length of the collimator.}
\label{fig:rv error with div angle}
\end{figure}

As mentioned earlier in this section, temperature changes not only shift the position of the spectral lines by altering the spacing between the etalon mirrors, but also affect their stability due to thermo-mechanical effects on the housing of the FP etalon wavelength calibrator. One can find more details on this regard in \cite{cersullo_2017_fp_spirou}. Temperature fluctuations can alter the collimator’s focal length, causing the input fiber to become slightly defocused and appear to have a larger effective diameter. These variation in the collimation optics directly affect the incident angle ($\theta$ in Equation \ref{eq:fp-interference}) of the light beam on the FP etalon. Under the small-angle approximation, we can derive the RV variation due to decentering and divergence angle:
\begin{equation}
\dfrac{\Delta v}{c} = \dfrac{1}{2\cos \theta}\left(\theta^2 + (\theta + d\theta)^2\right) \label{eq:rv error:divergence and decentering angle}
\end{equation}
Here, $\theta$ denotes the decentering angle ($\theta \approx \dfrac{\Delta}{f}$, here $\Delta$ is the decentering distance and $f$ is the focal length of the collimator lens) and $d\theta$ represents the divergence angle ($d\theta \approx \dfrac{d}{f}$, here $d$ is the diameter of the fiber tip). From this equation we can see that if $\theta \neq 0$, i.e there is decentering between the optical axis of the collimator and the etalon to start with then the effect of the variation in the divergence angle ($d\theta$) on the RV will be more aggressive compared to perfectly aligned ($\theta = 0$) system. 
This effect can be seen in Figure \ref{fig:rv error with div angle}. Here the right panel (varying decentering) shows a steeper RV variation than the left panel (varying divergence angle).

We can evaluate the sensitivity of the RV variation from the slope of the curves in Figure \ref{fig:rv error with div angle} or by derivating Equation \ref{eq:rv error:divergence and decentering angle} with respect to the divergence angle by a similar fashion shown in \cite{cersullo_2017_fp_spirou}. 
\begin{equation}
    \dfrac{d({\Delta v})}{d(d\theta)} = \dfrac{c}{\cos \theta} (\theta + d\theta).
    \label{eq:derivative div angle variation}
\end{equation}

In our case the change in focal length (for PAC076 lens; f = 125 mm) of the collimator with respect to temperature change is estimated as $\approx 3$ $\mu$m$/^\circ\mathrm{C}$ using the relation $\dfrac{df}{dT} = \alpha_\text{Al}f$ ($\alpha_{Al} = 2.3 \times 10^{-5}/^\circ\mathrm{C}$). Operating in an F/4 configuration, this leads to a corresponding apparent fiber diameter variation with temperature as $0.75~\mu$m$/^\circ\mathrm{C}$. This change in the apparent fiber size leads to change in the divergance angle which affects the RV stability. For a fiber size of 50 $\mu$m with a collimator lens of 125 mm focal length, we can estimate the RV variation using Equation \ref{eq:derivative div angle variation}. It comes out to be approximately 5 cm/s for a temperature stability $0.01 ~^\circ\mathrm{C}$.

Furthermore, the stability of the FP system is also influenced by the uniformity of illumination; even $\pm 0.5\%$ intensity fluctuations can introduce RV variations of up to 1 m/s as discussed in Hao (2021)\cite{Hao_2021_illumination_FP}. The study has suggested several approaches to mitigate the impact of non-uniform illumination, such as use of octagonal fibers and a double scrambler at the FP input. Therefore we are using a 50 $\mu$m octagonal fiber atthe FP input which has been shown to provide excellent scrambling performance\cite{Chazels_2010_octagonal_fiber} and in near future, we will also install a double scrambler at the FP input in conjunction with the octagonal fiber to further improve illumination stability.


\section{Installation, stability test and the initial result of the FP wavelength calibrator}

\begin{figure}[b]
\centering
\includegraphics[width=0.99\linewidth]{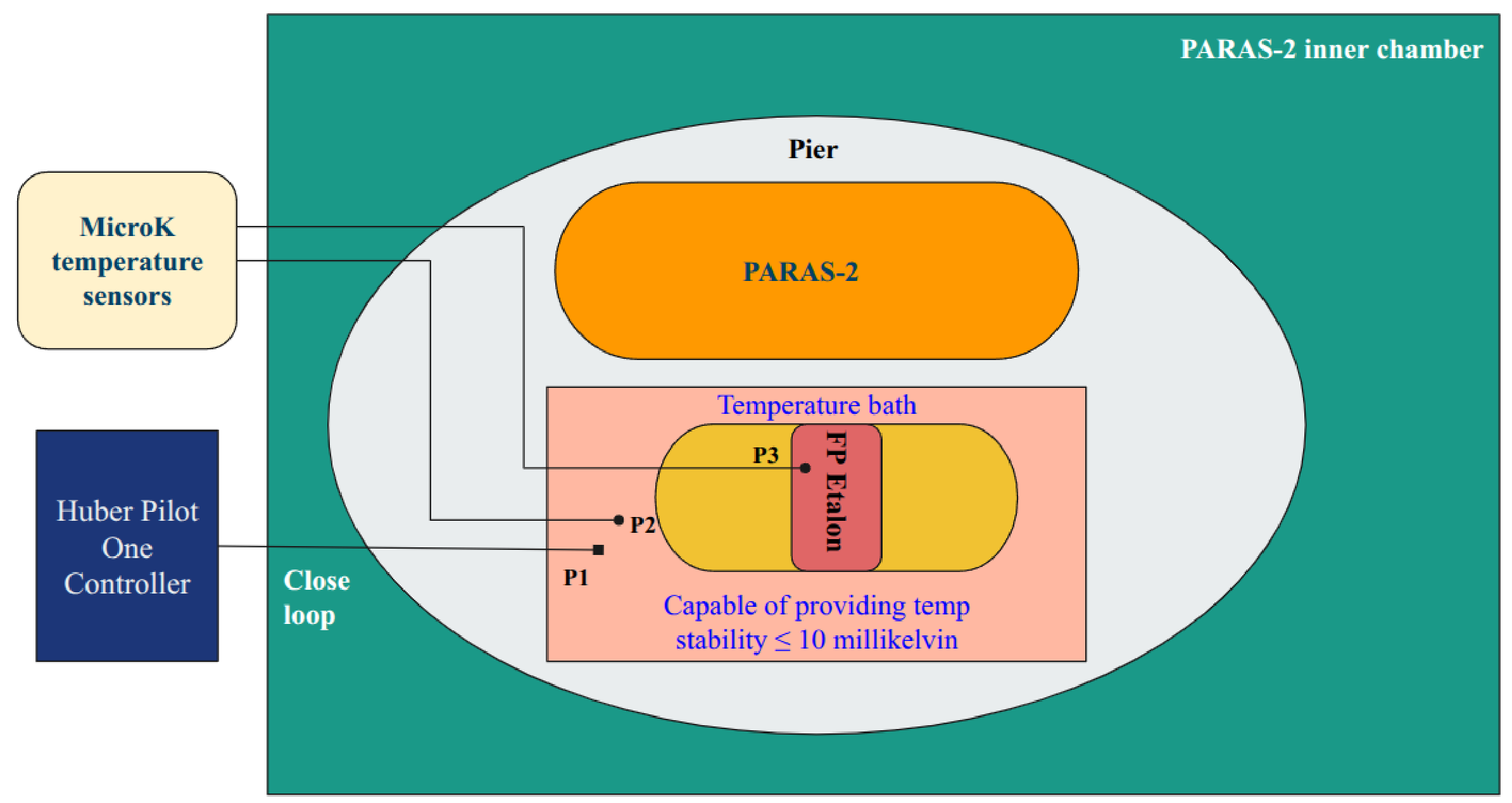}
\caption{The schematic of the FP system inside the temperature bath, controlled by the Huber Pilot ONE via a closed loop using the P1 sensor. The temperatures of both the bath and the FP optics are monitored using MicroK temperature sensors (P2 and P3). P2 sensor is inserted inside the water bath to re-measure the temperature control of the water bath. Whereas the P3 sensor is attached with the FP etalon optics. This image is not to scale.}
\label{fig:temperature schematic}
\end{figure}

The FP assembly is placed inside the inner chamber of the PARAS-2 (For more details please refer to Chakraborty (2018)\cite{Abhijit_2018_paras2}. However, for a schematic overview of the inner chamber of PARAS-2, please refer to Figure \ref{fig:temperature schematic}) to maintain the temperature stability of $0.01~^\circ\mathrm{C}$. The light from the FP etalon wavelength calibrator is fed to the spectrograph via the \textbf{CA}ssegrain \textbf{M}odule for \textbf{PA}RAS-2 \textbf{S}pectrograph\\(CAMPAS; for more details see \cite{kevi_2025_CAMPAS}). We are using a 365 $\mu$m circular fiber connecting the FP and the CAMPAS. There are provisions in the CAMPAS to feed the light from the FP to either Star fiber (75 $\mu$m) or Calibration fiber (75 $\mu$m) or into both at a time depending on observational requirements (for details one can refer to \cite{Abhijit_2024_bina_uar} and \cite{kevi_2025_CAMPAS}). CAMPAS also hosts additional calibration sources, including a tungsten lamp for flat-fielding and a uranium–argon hollow cathode lamp (UAr HCL)

\subsection{Temperature and pressure stability test results}\label{subsec:temp_press_stability}
We have shown that to achieve a RV stability better than 10 cm/s, we need to keep our system in a stable environment with temperature stability and pressure stability of $\leq 0.01 ~^\circ\mathrm{C}$  and $1\times 10^{-3}$ mbar respectively. 

\begin{figure}
    \centering
    \includegraphics[width=0.99\linewidth]{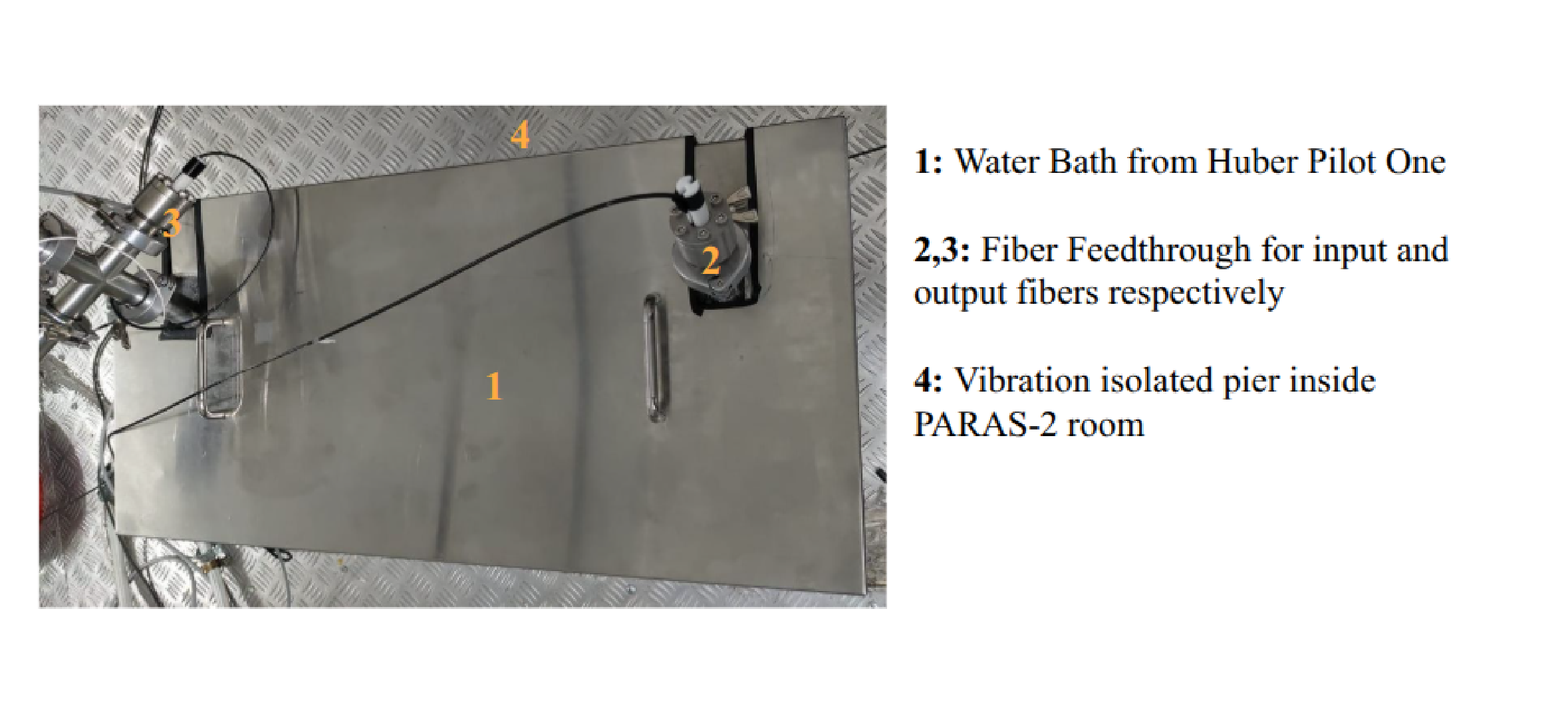}
    \caption{The FP vacuum enclosure (Figure \ref{fig:fpe real} (lower panel)) inside the water bath which is kept on pier inside the inner chamber (refer to Figure~\ref{fig:temperature schematic}).}
    \label{fig:fp inside bath}
\end{figure}

The temperature stability is achieved in two stages. First, the vacuum enclosure (Figure \ref{fig:fpe real} (lower panel)) is submerged in a temperature-controlled water bath (Figure \ref{fig:fp inside bath}) operated by the Huber Pilot ONE system. This setup allows active temperature regulation with a precision of $0.01 ~^\circ\mathrm{C}$ at $24 ~^\circ\mathrm{C}$. Second, this system is placed inside the inner chamber of the PARAS-2 spectrograph which itself have a good temperature stability with PID control which assist the Huber temperature control system to maintain a stability in RMS of $\Delta T \leq 0.01 ~^\circ\mathrm{C}$ . Under a stable environmental conditions of the PARAS-2 inner chamber, the FP system shows enhanced RMS variation in temperature of $0.002 ~^\circ\mathrm{C}$ over a 12-hour period.

We maintain the set temperature of the FP system at $24 ~^\circ\mathrm{C}$, which corresponds to the current ambient temperature of the PARAS-2 inner chamber (for details see \cite{Abhijit_2018_paras2,Abhijit_2024_bina_uar}). The entire setup is mounted on the pier of the PARAS-2 spectrograph to isolate it from external vibrations. The complete schematic of the system is shown in the Figure \ref{fig:temperature schematic}.

\begin{figure}
    \centering
    \includegraphics[width=0.9\linewidth]{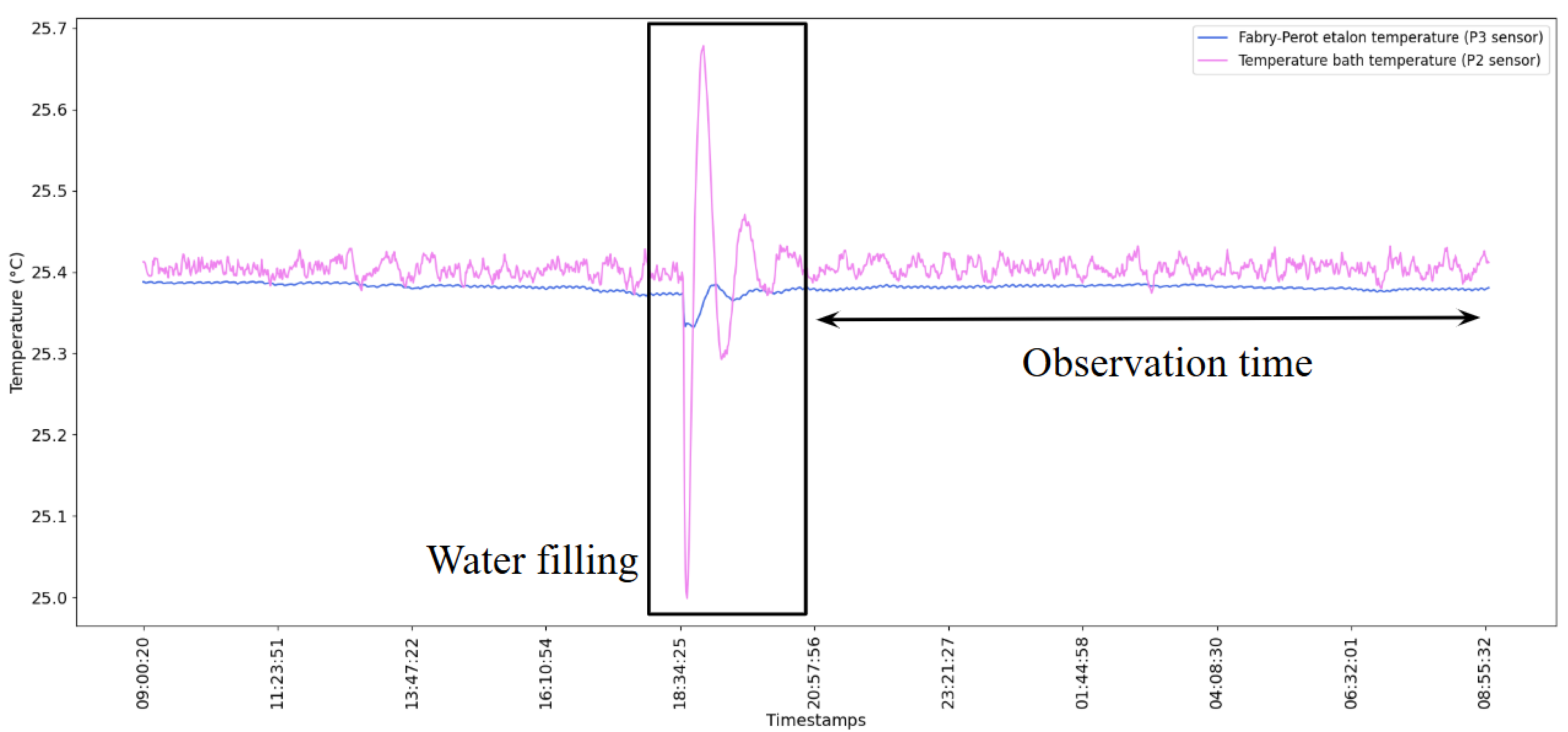}
    \caption{The temperature stability data measured by the MicroK sensors on 20 March 2025 is shown here. The significant variation observed around 19:00 hrs corresponds to the daily water refilling of the temperature bath. The blue curve represents the temperature measured at the vicinity to FP etalon optics (P3 sensors refer to Figure \ref{fig:temperature schematic}), showing that, once stabilised, the system maintains a RMS varition of approximately $0.002 ~^\circ\mathrm{C}$.}
    \label{fig:temperature stability}
\end{figure}

The P1 sensor shown in Figure~\ref{fig:temperature schematic} is responsible for active temperature control with the Huber Pilot One temperature controller. For higher precision measurement, two MicroK temperature sensors (PT100) mentioned as P2 and P3 are used to monitor temperature variations at the $\mu$K level. The water bath is refilled daily at approximately 19:00 hrs and it takes typically 2–3 hours to reach thermal equilibrium. We have presented the temperature variation data for approximately 12-hours window in Figure \ref{fig:temperature stability}. The plot shows that the temperature stability of approximately $0.002 ~^\circ\mathrm{C}$ during the observation time. These measurements are corresponds the temperature sensor placed near the FP etalon marked as P3 in Figure \ref{fig:temperature schematic}. The region marked as black rectangular box is the disturbance during refilling of the water bath.

To achieve a pressure stability of $1 \times 10^{-3}$ mbar, we perform continuous pumping with a turbo molecular pump (TMP). This operation help us to maintain a pressure stability of the order of $1\times10^{-4}$ mbar. The TMP is attached to the vacuum enclosure of the FP system using a metallic hose bellow of 2m in length alongside a vacuum valve. We have also attached a vibration damper to reduce the vibration generated by the continuous operation of the pump.

\begin{figure}
    \centering
    \includegraphics[width=0.9\linewidth]{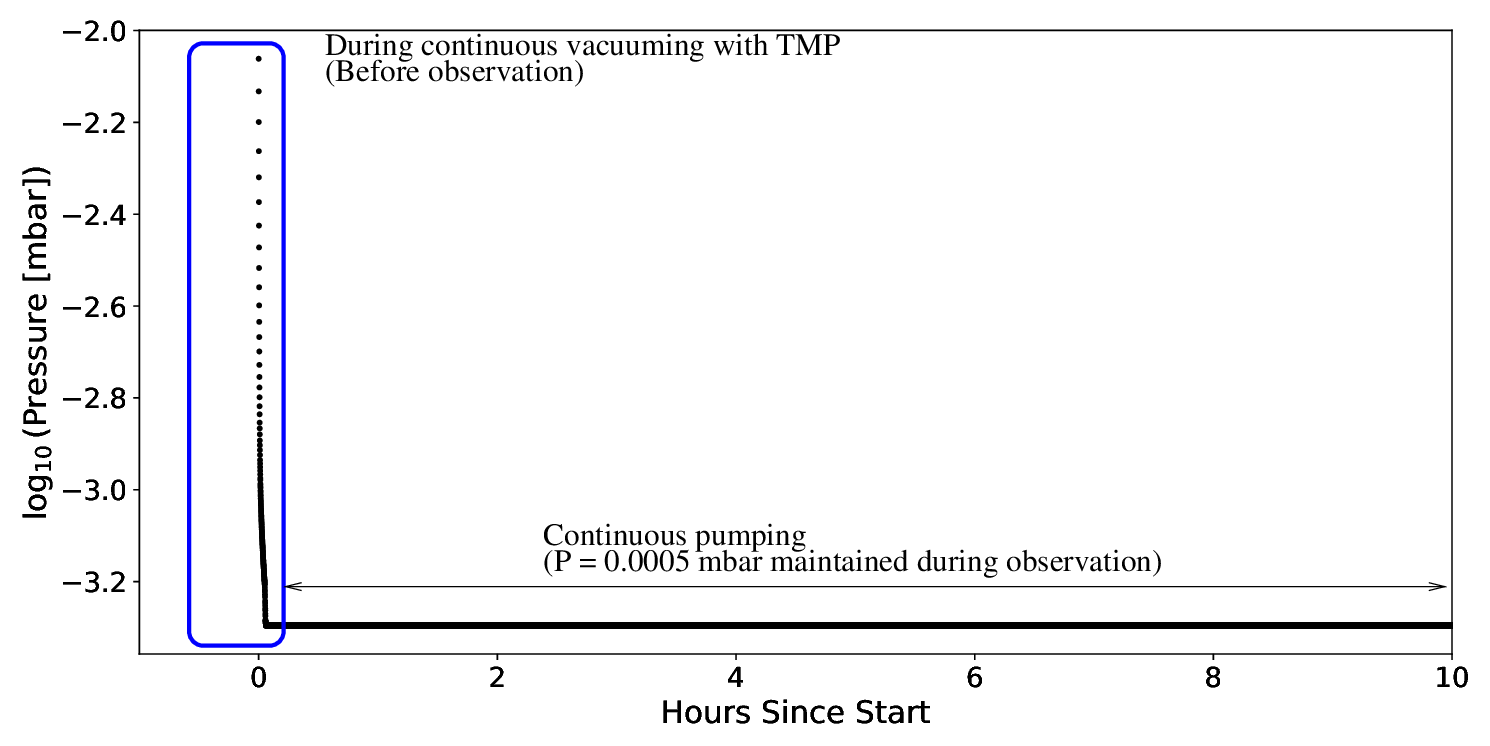}
    \caption{The vacuum stability data measured from just before starting the observation till the observation on 20 March 2025. During the operation, the pressure inside the vacuum chamber of FP remains stable at the $5\times10^{-4}$ mbar.}
    \label{fig:vacuum stability}
\end{figure}

We have shown in Figure \ref{fig:vacuum stability}, during the observation, pressure values shows a RMS variation within $5\times10^{-4}$ mbar, which is within the required pressure stability value.

\subsection{Initial results}

\subsubsection{Finesse measurement}
\begin{figure}[b]
    \centering
    \includegraphics[width=0.9\linewidth]{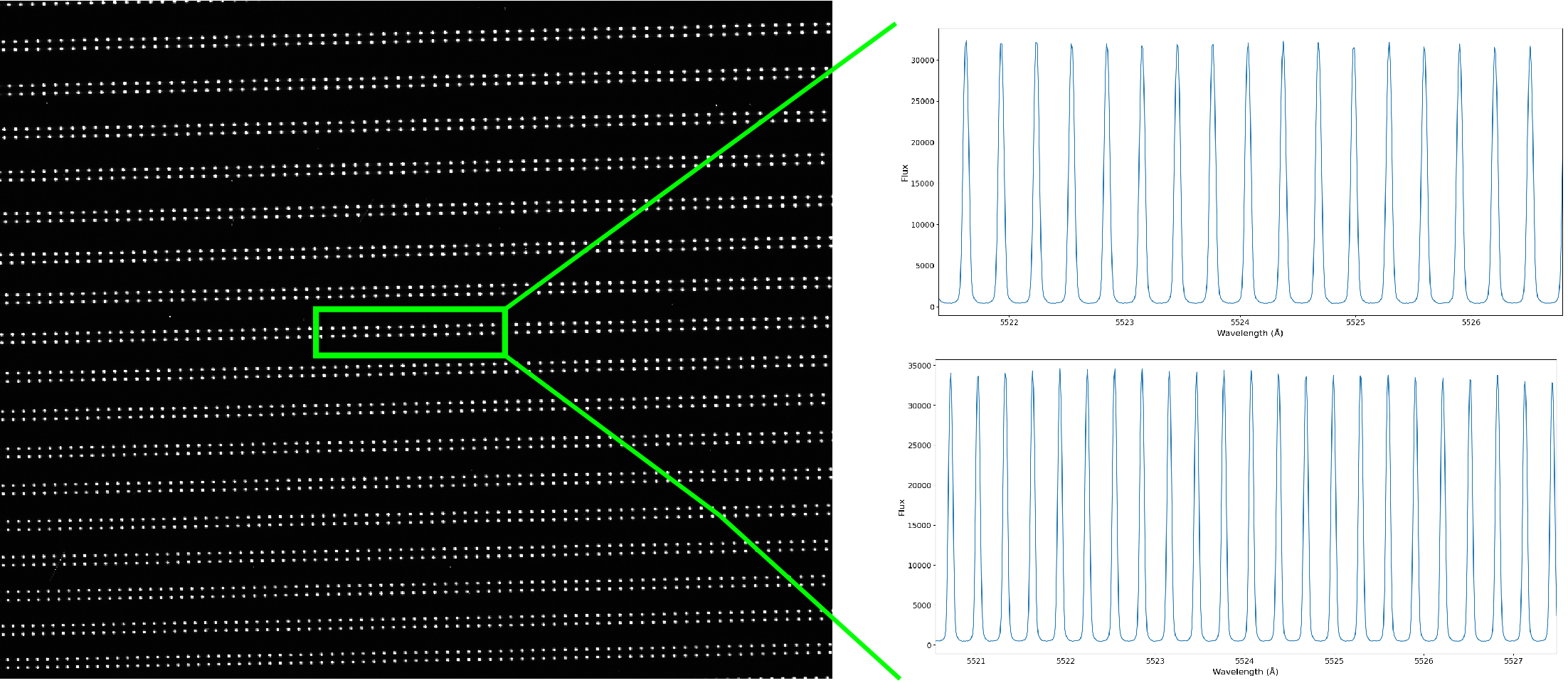}
    \caption{A snippet of the FP spectra taken by PARAS-2 spectrograph by illuminating both the star fiber and the calibration fiber. In the left panel we can see the CCD recorded image where as the right panel displays the flux vs wavelength for both star and calibration fiber near 5500 \AA.}
    \label{fig:real spectra}
\end{figure}

\begin{figure}[H]
    \centering
    \includegraphics[width=0.9\linewidth]{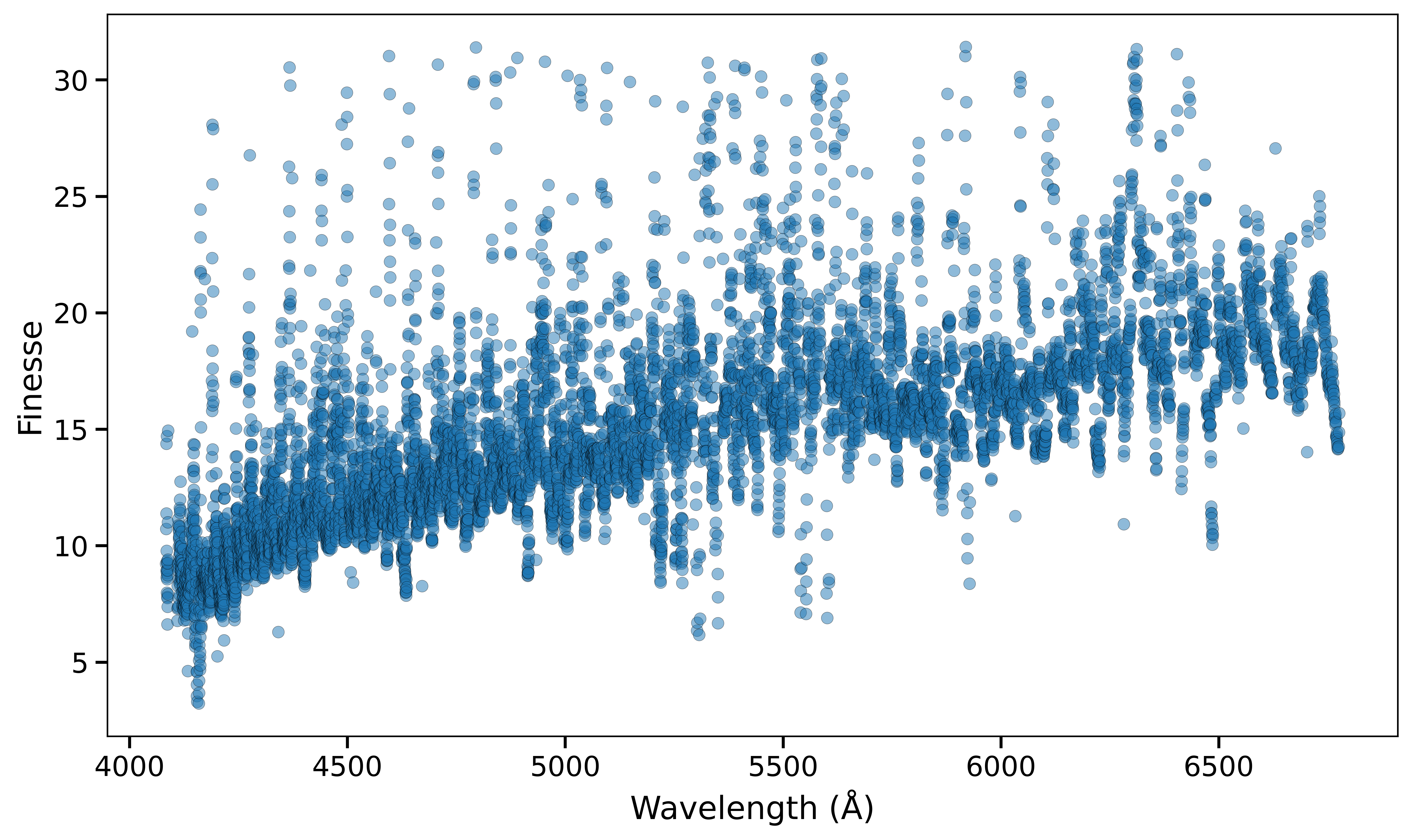}
    \caption{Finesse variation with wavelength measured using a 50 $\mu$m octagonal fiber at the input of the Fabry–Perot calibrator}
    \label{fig:finesse}
\end{figure}

We have identified over 10,000 FP spectral lines distributed across all 62 echelle orders between 4000 \AA~to 6900 \AA~of the PARAS-2 spectrograph. The left panel of Figure \ref{fig:real spectra} shows a section of the CCD-recorded raw image, where the FP lines appear uniformly spaced. The free spectral range (FSR) of these lines varies from approximately 0.16 \AA~in the blue (4000 \AA) to 0.49 \AA~in the red (7000 \AA), with a value of 0.3 \AA~near 5500 \AA, as illustrated in the right panel of Figure \ref{fig:real spectra}. With the 50 $\mu$m at the input of the FP, our obtained FWHM of the FP lines near 5500 \AA~is approximately 0.054 \AA. After correcting for the instrumental profile (IP) of PARAS-2—approximately 0.050 \AA~at this wavelength (corresponding to a resolving power R$\approx$1,10,000)—the finesse of the FP system in this region is estimated to be around 17, at 6900 \AA~it is estimated to be 19 which is higher than the targeted value of 10 (see \S \ref{sec:design consideration}). Accordingly at 4000 \AA~the observed finesse value is found to be close to 9. In Figure \ref{fig:finesse}, a linear trend between the finesse and wavelength is evident. This behaviour arises from the linear dependence of the divergence finesse on wavelength \cite{cersullo_2017_fp_spirou}. Since the divergence finesse sets the limit for our FP calibrator, this linear trend naturally appears in our measured finesse versus wavelength relation. The large scatter in finesse at the blue end of the spectrum is associated with low SNR, while additional scattered points are attributed to the FP lines located near the edges of the echelle orders.

\subsubsection{Initial radial velocity performance }

\begin{figure}[H]
    \centering
    \includegraphics[width=0.9\linewidth]{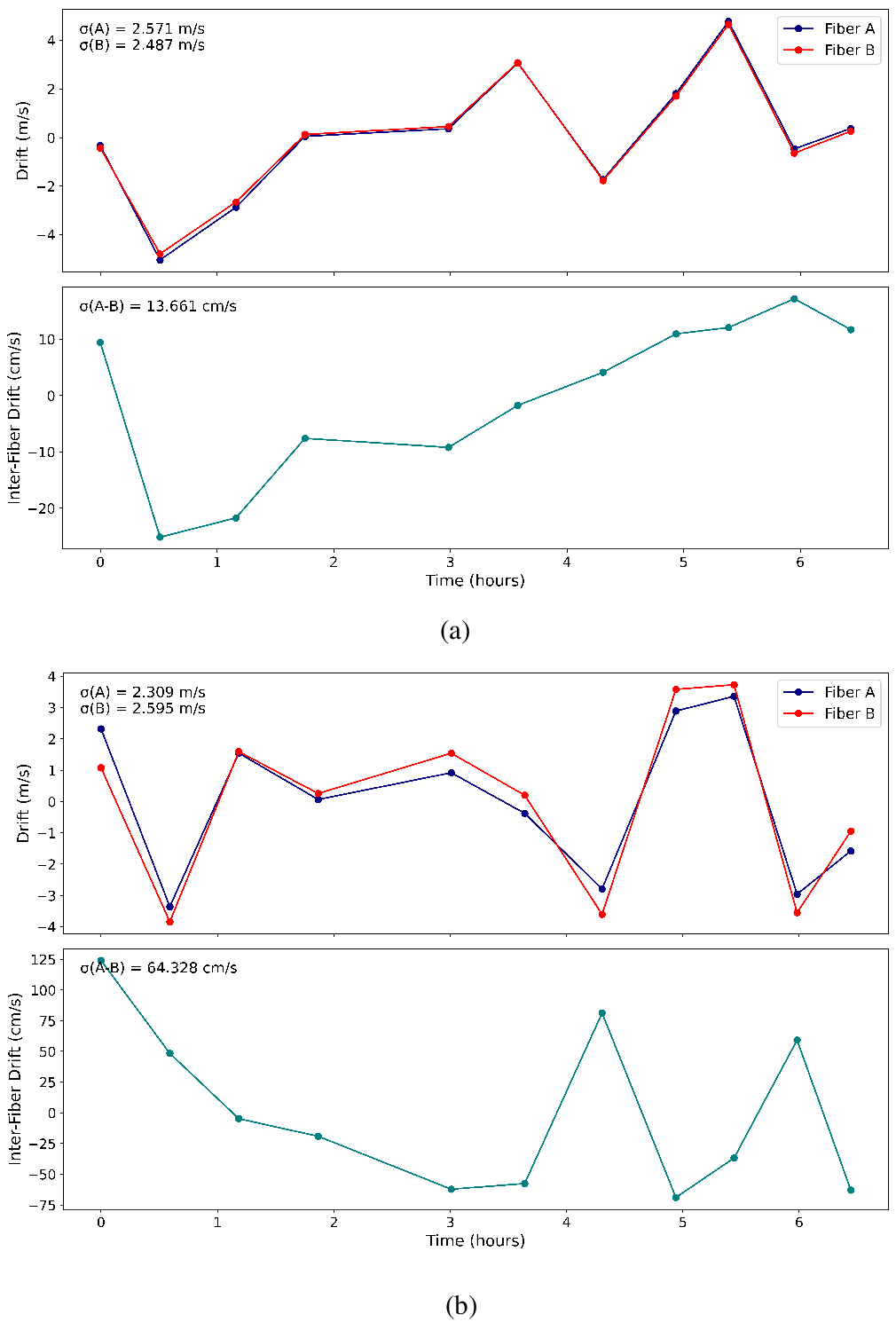}
    \caption{Instrumental drift measured with PARAS-2 using UAr (upper panel) and the FP calibrator (lower panel). Both calibration sources show comparable absolute drift variations ($\sim$ 2.5 m/s) in fiber A (star fiber) and fiber B (calibration fiber). However, the inter-fiber drift is higher for the FP ( $\sim$64 cm/s) compared to UAr ($\sim$14 cm/s).}
    \label{fig:rv stability}
\end{figure}

With the FP spectra recorded using PARAS-2, we performed an initial assessment of the radial-velocity stability of the FP wavelength calibrator and compared its performance with that of the UAr hollow cathode lamp (HCL). The UAr HCL is a well-established and highly stable reference source (see \cite{Abhijit_2024_bina_uar} for details). A more robust evaluation of the FP stability would involve simultaneous illumination of one fiber with a stable reference source (UAr in this case) and the other fiber with FP light, followed by a time-series sequence of exposures \cite{wildi_2010_fp}. However, the current PARAS-2 Cassegrain unit does not support simultaneous feeding of the two fibers with different calibration sources (see \cite{kevi_2025_CAMPAS} for a detailed description of the Cassegrain unit).
Given this limitation, we obtained alternating sequences of UAr–UAr (fiber A–fiber B) and FP–FP exposures, with one set acquired every 30 minutes over a duration of 6–8 hours. The instrumental drift was measured independently for both the UAr and FP spectra. Figure \ref{fig:rv stability} shows the measured drift over a $\sim$7-hour interval, with the left panel displaying the UAr drift and the right panel showing the FP drift. The absolute drift dispersion is comparable for both calibration sources and both fibers, with typical values in the range of 2–5 m/s and an observed dispersion of about 2.5 m/s in the shown figure.
However, in the inter-fiber drift behavior we can see a difference. For the UAr exposures, the inter-fiber drift variation is typically in the range of 10–15 cm/s. In contrast, the FP exposures exhibit a higher inter-fiber drift, with typical values of 40–70 cm/s. This may arise from spectral or illumination instabilities introduced by the Xe-arc lamp used to feed the FP, such as beam wandering, and/or from modal noise associated with the relatively large-core output fiber of the FP system.
We are actively investigating these effects and are working toward mitigating them by replacing the Xe-arc lamp with a laser-driven light source (LDLS) to reduce beam-wandering effects, as well as by testing smaller-core octagonal output fibers for the FP to improve modal stability and scrambling. Addressing these issues is currently a priority as we work to reduce the elevated inter-fiber drift observed with the FP calibrator. 

\section{Summary and Future work}
We have developed a FP wavelength calibrator along with a white light source for integration with the PARAS-2 spectrograph, which is attached with the PRL 2.5m telescope at the Mount Abu Observatory. All optical elements employed in these systems were procured off-the-shelf, while the optomechanical components were designed and fabricated in-house. The FP wavelength calibrator was developed considering the following technical and environmental specifications: (a) A spacer length of 5 mm made from Corning ULE glass, yielding a FSR of 0.3 \AA~at the central wavelength of 5500~\AA; (b) Finesse value ranges from 9 (near 4000 \AA) to 19 (near 6900 \AA) and at the central wavelength of 5500 \AA~it is approximately 17 with the 50 $\mu$m core octagonal fiber; (c) RMS of the temperature variation of $0.002 ~^\circ\mathrm{C}$  at a setpoint of $24 ~^\circ\mathrm{C}$, satisfying the design requirement of $0.01 ~^\circ\mathrm{C}$; and (d)RMS of the pressure variation within $5\times10^{-4}$ mbar is maintained under continuous vacuum pumping during observations. To mitigate vibrations induced by the pumping process, a vibration damper is employed.  Based on these RMS values of temperature and pressure, the theoretical RV variation of the FP wavelength calibrator as mentioned in \S\ref{sec:stability} and Figure \ref{fig:fp_temp_variation} is expected to remain within 10 cm/s, thereby meeting our target stability requirement. Our comparison of the RV stability of the FP with that of the UAr source shows a similar level of absolute drift's variation in both fibers (2–5 m/s). However, the inter-fiber drift measured with the FP (40–70 cm/s) is larger than that obtained with the UAr source (10–15 cm/s).To further reduce the higher-than-expected inter-fiber drift observed in our current measurements, we outline the following planned upgrades and investigations for the near future:
\begin{itemize}
\item Integration of a double-scrambler module, followed by on-sky validation of the upgraded FP + double-scrambler system using RV standard stars to assess performance under real observing conditions.
\item Real-time evaluation of the intrinsic stability of the FP system, particularly its sensitivity to pressure and temperature variations.
\item Assessment of the existing Xe arc lamp and potential replacement with a laser-driven light source (LDLS) to achieve improved spectral stability and longer operational lifetime.
\item Investigation of the FP wavelength calibrator’s performance using a smaller-core output fiber to reduce modal noise and improve stability.
\end{itemize}

\subsection*{Disclosures}
The authors declare that there are no financial interests, commercial affiliations, or other potential conflicts of interest that could have influenced the objectivity of this research or the writing of this paper.

\subsection*{Code and Data Availability}
The code and data that support the findings of this article can be provided on reasonable request to the corresponding author.

\subsection* {Acknowledgments}
We acknowledge the support from PRL-DOS (Department of Space, Government of India) and thank the Director of PRL for funding the PARAS-2 spectrograph and the FP etalon wavelength calibrator project. We extend our sincere gratitude to all the staff of the PRL workshop for their assistance during the fabrication of the FP and Xe arc lamp housings. We also thank the staff at the PRL Mount Abu Observatory for their support during the system installation. We are grateful to Prof. François Wildi and Prof. Francesco Pepe for valuable discussions on the FP system design. We thank Dr. Maria Federica Cersullo for her suggestions regarding various simulations used in the system design. This paper utilized Python for generating plots using the Matplotlib library \cite{Hunter_2007_matplotlib} and different calculations using NumPy \cite{numpy}.


\section*{Biography}
Shubhendra Nath Das is a Senior Research Fellow (Ph.D. scholar) at the Physical Research Laboratory (PRL), Ahmedabad, India. He completed his Master of Science in Physics from the University of Calcutta. His doctoral work primarily focuses on astronomical instrumentation and exoplanetary studies. He has been responsible for the assembly, integration, and testing (AIT), as well as the off-sky and on-sky characterization, of a Fabry–Perot wavelength calibrator developed for the PARAS-2 high-resolution spectrograph (R $\sim$ 1,10,000) on the PRL 2.5-m telescope.
His research interests span optical instrumentation for ground-based telescopes, high-resolution spectroscopy, and exoplanet science, with a particular focus on the detection and characterization of small, terrestrial exoplanets.

\end{spacing}
\end{document}